\documentclass[letterpaper,twocolumn,10pt]{article}
\usepackage{usenix,epsfig,endnotes}
\usepackage{cite}
\usepackage{amsfonts}
\usepackage{xspace}
\usepackage{pifont}
\usepackage[usenames,dvipsnames]{color}
\usepackage{multirow}
\usepackage[english]{babel}
\usepackage{graphicx}
\usepackage{flushend}
\usepackage{authblk}
\usepackage{amsbsy}

\newtheorem{theorem}{Theorem}

\usepackage{wrapfig}
\usepackage{graphicx}
\usepackage{caption}

\usepackage{fancybox}
\usepackage{calc}
\usepackage{url}
\usepackage[hang,scriptsize,tight,nooneline]{subfigure}
\usepackage{booktabs}
\usepackage{times}
\usepackage{enumitem}

\newcommand{\pbgnew}[1]{{\color{Black}#1}}

\newcommand{\qingxic}[1]{{}}

\newcommand{\wxzc}[1]{{}}

\newcommand{\cut}[1]{}

\newcommand{\parheading}[1]{\medskip{} \noindent \textbf{#1}}

\newcommand{\name}{PCC\xspace}
\newcommand{\eps}{\varepsilon}

\begin{document}


\title{\vspace{-0.23in}PCC: Re-architecting Congestion Control for Consistent High Performance}

\author[*]{Mo Dong}
\author[*]{Qingxi Li}
\author[**]{Doron Zarchy}
\author[*]{P. Brighten Godfrey}
\author[**]{Michael Schapira}
\affil[*]{University of Illinois at Urbana-Champaign}
\affil[**]{Hebrew University of Jerusalem}


\maketitle

\begin{abstract}
\textnormal{TCP and its variants have suffered from surprisingly poor performance for decades. We argue the TCP family has little hope to achieve consistent high performance due to a fundamental architectural deficiency: hardwiring packet-level events to control responses without understanding the real performance result of its actions. We propose Performance-oriented Congestion Control (PCC), a new congestion control architecture in which each sender 
continuously observes the connection between its \emph{actions} and \emph{empirically experienced performance}, enabling it to consistently adopt actions that result in high performance.  We prove that PCC converges to a stable and fair equilibrium. Across many real-world and challenging environments, PCC shows consistent and often $10\times$ performance improvement, with better fairness and stability than TCP.  PCC requires no router hardware support or new packet format.}


\end{abstract}

\vspace{-5mm}
\section{Introduction}
\label{sec:intro}
\vspace{-2mm}

In the 26 years since its deployment, TCP's congestion control architecture has been notorious for degraded performance in numerous real-world scenarios. TCP performs poorly on lossy links, penalizes high-RTT flows, underutilizes high bandwidth-delay product~(BDP) connections, cannot handle rapidly changing networks, can collapse under data center incast~\cite{understandingincast} and incurs very high latency with bufferbloat~\cite{bufferbloat} in the network.

Solutions requiring in-network hardware or protocol changes~\cite{xcp,rcp} have rarely seen widespread deployment. More commonly, end-host-based protocol ``patches'' have addressed problems in specific network conditions such as high BDP links~\cite{wj+07, CUBIC}, satellite links~\cite{winds, hybla}, data center~\cite{ictcp, dctcp}, wireless and lossy links~\cite{liu2008tcp, westwood}, and more.  The fact that there are so many TCP variants suggests that each is only a point solution: they yield better performance under specific network conditions, but break in others. Worse, we found through real-world experiments that in many cases the \textbf{performance of these TCP variants is still quite far from optimal even in the network conditions towards which they are specially engineered}. For example: TCP CUBIC, optimized for high BDP links, commonly operates $10\times$ away from optimal throughput on the commercial Internet; TCP Hybla is optimized for lossy and long RTT satellite links, but in practice can barely get $6\%$ of capacity (\S\ref{sec:eval}). We found it surprising that moving data across the network, which so many applications depend on, still suffers from such degraded performance.

Thus, despite the large number of TCP variants, the fundamental problem remains largely unsolved: \textbf{achieving consistently high performance over complex real-world network conditions}.
We argue this is indeed a very difficult task within TCP's \emph{\textbf{hardwired mapping}} rate control architecture: hardwiring certain \emph{predefined packet-level events} to certain \emph{predefined control responses}. In a hardwired mapping, TCP reacts on packet-level events that can be as simple as ``one packet loss'' (TCP New Reno) or can involve multiple signals like ``one packet loss and RTT increased by $x\%$'' (TCP Illinois).  Similarly, the control response might be ``halve the rate'' (New Reno) or a more complex action like ``reduce the window size $w$ to $f(\Delta RTT)w$'' (Illinois).  The defining feature is that the control action is a \emph{deterministic hardwired function} of packet-level events.

The design rationale behind the hardwired mapping architecture is to make assumptions about the packet-level events. When it sees a packet-level event, TCP \emph{assumes} the network is in a certain state (e.g. congestion, queue building up) and  tries to optimize performance by triggering a predefined control behavior as the response to that \emph{assumed} state. In real networks, assumptions fail but TCP still mechanically carries out the mismatched control response, resulting in severely degraded performance. Take an event-control pair from textbook TCP: a packet loss halves the congestion window size. TCP assumes that the loss indicates congestion in the network. When the assumption is violated, halving the window size will cause severe performance degradation (e.g. if loss is random, rate should stay the same or increase).  It is fundamentally hard to formulate an ``always optimal'' hardwired mapping in a complex real-world network because the actual optimal response to an event like a loss (i.e. decrease rate or increase? by how much?) is highly sensitive to conditions including random loss, router buffer size, competing flows' RTT and so on.

Remy~\cite{remy, learnabilitytcp} (\S\ref{sec:rel}) pushes TCP's architecture perhaps as far as it can go, by simulating many possible TCP-like protocols to find which hardwired controls tend to perform well in a \emph{assumed network scenario}. But even for Remy's algorithmically-designed hardwired mapping, performance degrades~\cite{learnabilitytcp} when number of senders, RTT and number of bottlenecks in the real network deviate from assumed input parameters. Moreover, Remy ignores random loss in its input model.

In fact, today's production networks have moved to a level of complexity beyond the assumptions embedded in any hardwired mapping: old and unreliable routers, failing wires, links from Kbps to 100 Gbps, unstable routing paths, AQMs, software routers, rate shaping at gateways, virtualization layers and middleboxes like firewalls, packet inspectors and load balancers. All these factors add so much complexity that can almost surely violate any TCP-like hardwired mapping's relatively simplistic assumptions about networks. Most unfortunately, when a violation happens, TCP still rigidly carries out the harmful control action because \textbf{\emph{it does not see its control action's actual effect on performance}}.


We propose a new congestion control architecture: Performance-oriented Congestion Control~(PCC). PCC rises from where TCP fails, by associating a control action (change of sending rate) directly with its effect on real performance. For example, when a sender changes its rate to $r$ and gets SACKs after sending at this rate, instead of trigging any predefined control action, PCC aggregates these packet-level events into meaningful performance metrics (throughput, loss rate, latency, etc.) and combines them into a numerical value $u$ via a utility function describing objectives like ``high throughput and low loss rate''. With this capability of understanding the real performance result of a particular sending rate, PCC then directly observes and compares different sending rates' resulting utility and learns how to adjust its rate to improve empirical utility through a learning control algorithm. By avoiding any assumptions about the underlying potentially-complex network, PCC tracks the \emph{empirically} optimal sending rate and thus achieves consistent high performance. PCC's learning control is selfish in nature, but surprisingly, using a widely applicable utility function, competing PCC senders provably converge to a fair equilibrium. Indeed, experiments show PCC achieves similar convergence time to TCP with significantly smaller rate variance.

Moreover, as discussed in detail later (\S\ref{sec:diffutility} and \S\ref{sec:eval-fq}), PCC provides a \textbf{flexibility beyond TCP's architecture}: expressing different objectives of data transfer, e.g. $throughput/latency$, with different utility functions.

With handling real-world complexity as goal, we evaluated a PCC implementation \textbf{not by simulation but in large-scale and real-world networks}. \emph{With no tweak of its control algorithm}, PCC achieves consistent high performance and significantly beats \emph{specially engineered} TCPs on various network environments: \textbf{(a.)} the wild and complex global commercial Internet (often more than $\pmb{10\times}$ the throughput of TCP CUBIC); \textbf{(b.)} inter-data center networks ($\pmb{5.23\times}$ vs. TCP Illinois); \textbf{(c.)} emulated satellite Internet links ($\pmb{17\times}$ vs TCP Hybla); \textbf{(d.)} unreliable lossy links ($\pmb{10-37\times}$ vs Illinois); \textbf{(e.)} unequal RTT of competing senders (an \textbf{architectural cure} to RTT unfairness); \textbf{(f.)} shallow buffered bottleneck links (up to $\pmb{45\times}$ higher performance, or $\pmb{13\times}$ less buffer to reach $90\%$ throughput); \textbf{(g.)} rapidly changing networks ($\pmb{14\times}$ vs CUBIC, $\pmb{5.6\times}$ vs Illinois). PCC performs similar to ICTCP~\cite{ictcp} in \textbf{(h.)} incast scenario in data centers.

Though it is a substantial shift in architecture, PCC can be deployed by only replacing the sender-side rate control of TCP.  It can also deliver real data today with a user-space implementation: \url{https://github.com/modong/pcc}.

\vspace{-2mm}

\vspace{-4mm}
\section{PCC Architecture}
\vspace{-3mm}

\subsection{The Key Idea}
\label{sec:new_arch}
\vspace{-3mm}

Suppose flow $f$ is sending a stream of data at some rate and a packet is lost.  How should it react?  Should it slow the sending rate, or increase, and by how much? Or leave the rate unchanged?  This is a difficult question to answer because real networks are complex.  A single loss might be the result of \emph{many} possible underlying network scenarios. To pick a few:
\begin{itemize}
\vspace{-3mm}
	\item $f$ may be responsible for most of congestion.  Then, it should decrease its rate.
\vspace{-3mm}
	\item $f$ might traverse a shallow buffer on a high-BDP link, with the loss due to bad luck in statistical multiplexing rather than high link utilization.  Then, backing off a little is sufficient.
\vspace{-3mm}
    \item There may be a higher-rate competing flow. Then, $f$ should maintain its rate and let the other back off.
\vspace{-6mm}	
\item There may be random non-congestion loss somewhere along the path.  Then, $f$ should maintain or increase its rate.
\end{itemize}
\vspace{-2mm}

Clasically, TCP assumes a packet loss indicates non-negligible congestion, and that halving its rate will improve network conditions.  However, this assumption is false and will degrade performance in three of the four scenarios above.  Fundamentally, picking an optimal \emph{predefined and hardwired} control response is hard because for the same packet-level events, a control response optimal under one network scenario can decimate performance in even a slightly different scenario.  The approach taken by a large number of TCP variants is to design more sophisticated packet-level events and control actions; but this approach does not solve the fundamental problem, because \emph{they still hardwire predetermined events to predetermined control responses}, thus inevitably embedding unreliable assumptions about the network. The real network can have complexity beyond what any hardwired mapping can model, as discussed in~\S\ref{sec:intro}, and when the unreliable assumptions are violated, performance degrades severely. For example, TCP Illinois~\cite{liu2008tcp} uses both loss and delay to form a complicated packet-level event-control mapping, but its throughput catastrophically collapses with even a tiny amount of random loss, or when the network is dynamically changing (\S\ref{sec:eval}). More examples are in \S\ref{sec:rel}.


Most unfortunately, if some control actions are indeed harming performance, TCP can still mechanically ``jump off the cliff'' again and again, because it does not notice the control action's actual effect on performance.


But that observation points toward a solution.  Can we design a control algorithm that directly understands whether or not its actions actually improve performance?

\begin{figure}[t]
\label{fig:PCCvsTCP}
\centering
{{\includegraphics[width=3in]{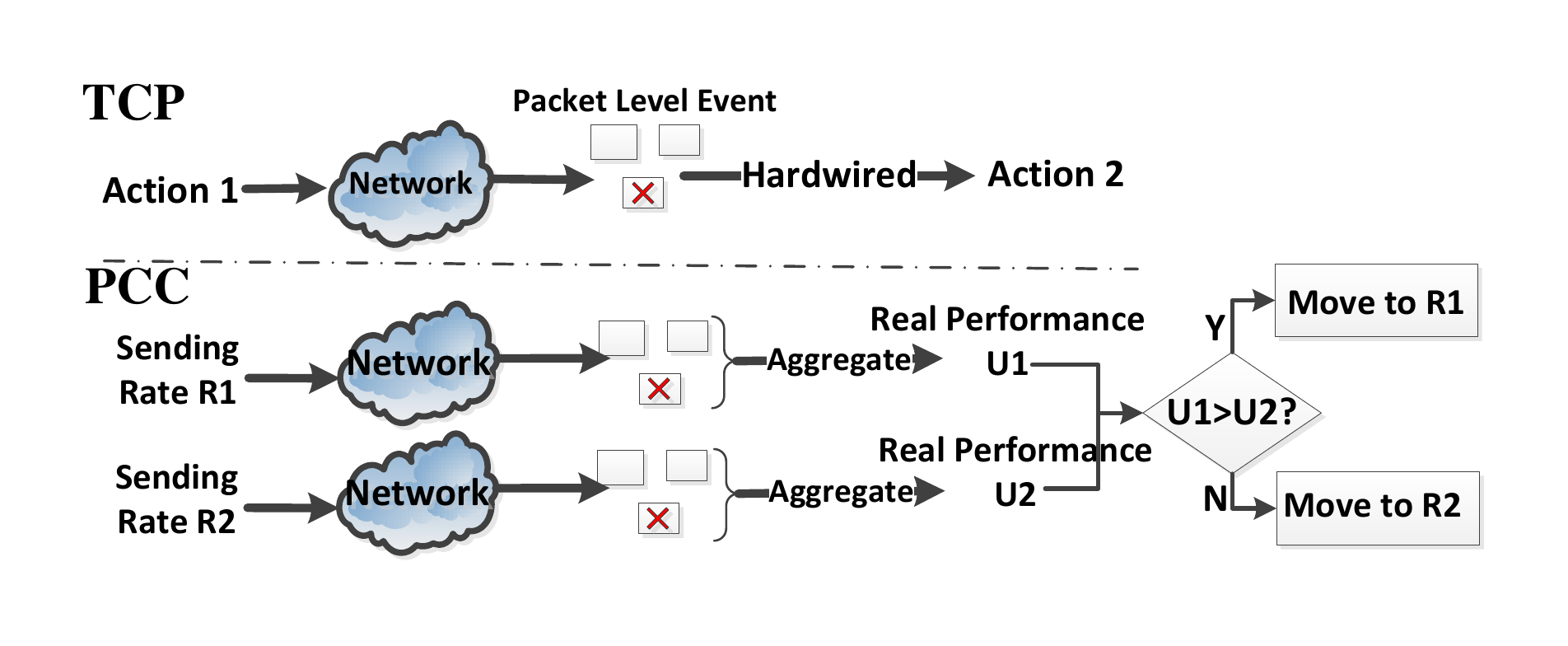}}
\vspace{-2mm}
\caption{\small \em \label{bbccvstcp}
The decision-making structure of TCP and PCC.}
}
\vspace{-5mm}
\end{figure}

Conceptually, no matter how complex the network is, if a sender can directly measure that rate $r_1$ results in better performance than rate $r_2$, it has some evidence that $r_1$ is better than sending at $r_2$ --- at least for this one sender. This example illustrates the key design rationale behind \textbf{Performance-oriented Congestion Control} (PCC): we make control decisions based on \emph{empirical evidence pairing \textbf{actions} with directly \textbf{observed performance} results}.

PCC's control action is its choice of sending rate. PCC divides time into continuous equal time periods, called \emph{monitor intervals}, whose length is normally one to two RTTs. In each monitor interval, PCC tests an action: it picks a sending rate, say $r$, and sends data at this rate through the interval. After about an RTT, the sender will see selective ACKs (SACK) from the receiver, just like TCP. However, it does not trigger any predefined control response. Instead, PCC aggregates these SACKs into meaningful performance metrics including throughput, loss rate and latency. These performance metrics are combined to a numerical utility value, say $u$, via a \emph{utility function}. The utility function can be customized for different data transmission objectives, but for now the reader can assume the objective of ``high throughput and low loss rate'', such as $u=T-L$ (where $T=$ throughput and $L=$ loss rate) which will capture the main insights of PCC. The end result is that PCC knows when it sent at rate $r$, it got utility of $u$.

The preceding describes a single ``experiment'' through which PCC associates a specific \emph{action} with an observed \emph{resulting utility}.  PCC runs these experiments continuously, comparing the utility of different sending rates so it can track the optimal action over time. More specifically, PCC runs a gradient-ascent online learning algorithm. When starting at rate $r$, it tests rate $(1+\eps)r$ and rate $(1-\eps)r$, and moves in the direction (higher or lower rate) that empirically results in higher utility.  It then continues in this direction as long as utility continues increasing.  If utility falls, it returns to a decision-making state where it again tests both higher and lower rates to determine which produces higher utility.

Note that PCC does not send occasional probes or use throwaway data for measurements.  It observes the results of its actual control decisions on the application's real data and does not pause sending to wait for performance result. 

\vspace{-2mm}
\parheading{We now return to the example} of the beginning of this section.  Suppose PCC is testing rate $100$~Mbps in a particular interval, and will test $105$ Mbps in the following interval.  If it encounters a packet loss in the first interval, will PCC increase or decrease?  In fact, there is no specific event in a single interval that will always cause PCC to increase or decrease its rate.  Instead, PCC will calculate the utility value for each of these two intervals, and move in the direction of higher utility.  For example:
\begin{itemize}
\vspace{-3mm}
	\item If the network is congested as a result of this flow, then it is likely that sending at $100$~Mbps will have similar throughput and lower loss rate, resulting in higher utility. PCC will decrease its rate.
\vspace{-3mm}
	\item If the network is experiencing random loss, PCC is likely to find that the period with rate $105$~Mbps has similar loss rate and slightly higher throughput, resulting in higher utility.  PCC will therefore increase its rate despite the packet loss.
\vspace{-3mm}
\end{itemize}

Throughout this process, PCC makes no assumptions about the underlying network conditions. It treats the network as a black box, observing which actions empirically produce higher utility and therefore achieving consistent high performance.

\vspace{-2mm}
\parheading{Decisions with noisy measurements.}  PCC's experiments on the live network will tend to move its rate in the direction that empirically improves utility.  But it may also make some incorrect decisions.  In the example above, if the loss is random non-congestion loss, it may randomly occur that loss is substantially higher when PCC tests rate $105$~Mbps, causing it to pick the lower rate.  Alternately, if the loss is primarily due to congestion from this sender, unpredictable external events (perhaps another sender arriving with a large initial rate while PCC is testing rate $100$ Mbps) might cause rate $105$~Mbps to have higher throughput and lower loss rate.  More generally, the network might be changing over time for reasons unrelated to the sender's action.  This adds noise to the decision process: PCC will on average move in the right direction, but may make some unlucky errors.

We improve PCC's decisions with \textbf{multiple randomized controlled trials (RCTs)}. Rather than running two tests (one each at $100$ and $105$~Mbps), we conduct four in randomized order---e.g. perhaps $(100,105,105,100)$.  PCC only picks a particular rate as the winner if utility is higher in \emph{both} trials with that rate.  This produces increased confidence in a causal connection between PCC's action and the observed utility.  If results are inconclusive, so each rate ``wins'' in one test, then PCC maintains its current rate, and we may have reached a local optimum (details follow later).

As we will see, without RCTs, PCC already offers a dramatic improvement in performance and stability compared with TCP, but RCTs further reduce rate variance by up to $65$\%.  Although it might seem that RCTs will double convergence time, this is not the case because they let PCC make \emph{better} decisions; overall, RCTs improve the stability/convergence-speed tradeoff space.

\vspace{-2mm}
\parheading{Many issues remain.}  We next delve into fairness, convergence, and choice of utility function; deployment; and flesh out the mechanism sketched above.

\vspace{-4mm}
\subsection{Fairness and Convergence}
\vspace{-2mm}
\label{sec:safety}

Each PCC sender optimizes its utility function value based only on locally observed performance metrics. Therefore, PCC's control is selfish. However, local selfishness does not indicate loss of global stability, convergence and fairness. We outline the proof (due to space limit) that with a certain kind of ``safe'' utility function and a simple control algorithm, selfish senders will voluntarily  converge to fair rate equilibrium.

We assume $n$ PCC senders $1,\ldots,n$ send traffic across a bottleneck link of capacity $C>0$. Each sender $i$ chooses a sending rate to optimize its utility function $u_i$. We choose a utility function expressing the common application-level goal of ``high throughput and low loss'':
\vspace{-0.4mm}$$u_i(x)=T_i(x)\cdot Sigmoid(L(x) - 0.05)-x_i\cdot L(x)$$\vspace{-0.2mm}where $x=(x_1,\ldots,x_n)$ is a global state of sending rates, $L(x)=\max\{0,1-\frac{C}{\Sigma_j x_j}\}$ is the per-packet loss probability, $T_i(x)=x_i(1-L(x))$ is sender $i$'s throughput, and
$Sigmoid(y)=\frac{1}{1+e^{\alpha y}}$ for some $\alpha >0$, to be chosen later.

The above utility function is derived from a simpler starting point: $u_i(x)=T_i(x)-x_i\cdot L(x)$, i.e., $i$'s throughput minus its packet loss rate. However, this utility function will make loss rate approach $50\%$ when the number of competing senders increases. Therefore, we include the sigmoid function as a ``cut-off''. When $\alpha$ is ``big enough'', i.e., $L(x) > 0.05$, $Sigmoid(L(x)-0.05)$ will rapidly get closer to $0$, leading to a negative utility for the sender. Thus, we are setting a barrier that caps the overall loss rate at about 5\% \pbgnew{in the worst case}.

\vspace{-3mm}
\begin{theorem}\label{theorem1}
	When $\alpha\geq \max\{2.2(n-1),100\}$, there exists a unique stable state of sending rates $x^*_1,\ldots,x^*_n$ and, moreover, this state is fair, i.e., $x^*_1=x^*_2=\ldots=x^*_n$.
\end{theorem}
\vspace{-3mm}




To prove Theorem ~\ref{theorem1}, we first prove that $\Sigma_j x_j$ will always be restricted to the region of $(C,\frac{20C}{19})$. Under this condition, our setting can be formulated as a concave game~\cite{Rosen1965}. This enables us to use properties of such games to conclude that a unique rate equilibrium exists and is fair, i.e. $x^*_1=x^*_2=\ldots=x^*_n$. (Full proof: \cite{pccfullproof})

Next, we show that a simple control algorithm can converge to that equilibrium. At each time step $t$, each sender $j$ updates its sending rate according to $x^{t+1}_j =x^t_j(1+\eps)$ if $u_j(x^t_j(1+\eps),x^t_{-j})>u_j(x^t_j(1-\eps),x^t_{-j})$, and $x^{t+1}_j =x^t_j(1-\eps)$ otherwise.  Here $x_{-j}$ denotes the vector of sending rates of all senders except for $j$, and $\eps>0$ is a small number ($\eps=0.01$, in the experiment). In this analysis, senders concurrently update their rates, but each sender decides based on a utility comparison as if it were the only one changing.  We believe this is a reasonable simplified model for analysis, coupled with experimental evidence.  (We also conjecture the model can be relaxed to allow for asynchrony.)  We discuss in~\S\ref{sec:prototype} our implementation with practical optimizations of the control algorithm.

\vspace{-3mm}
\begin{theorem}\label{ConvergenceTheorem}
If all senders follow the above control algorithm, for every sender $j$, $x_j$ converges to the domain $(\hat{x}(1-\eps)^2,\hat{x}(1+\eps)^2)$, where $\hat{x}$ denotes the sending rate in the unique stable state.
\end{theorem}
\vspace{-3mm}

It might seem surprising that PCC uses \emph{multiplicative} rate increase and decrease, yet achieves convergence and fairness. If TCP used MIMD, in an idealized network senders would often get the same back-off signal at the same time, and so would take the \emph{same multiplicative decisions in lockstep}, with the ratio of their rates never changing.  In PCC, senders make \emph{different} decisions.  Consider a 100 Mbps link with sender $A$ at rate 90 Mbps and $B$ at 10 Mbps.  When $A$ experiments with slightly higher and lower rates $(1\pm \eps)90$ Mbps, it will find that it should decrease its rate to get higher utility because when it sends at higher than equilibrium rate, the loss rate dominates the utility function.  However, when $B$ experiments with $(1\pm \eps)10$ it finds that loss rate increase is negligible compared with its improved throughput.  This occurs precisely because $B$ is responsible for little of the congestion.  In fact, this reasoning (and the formal proof of the game dynamics) is \emph{independent of the step size} that the flows use in their experiments: PCC senders move towards the convergence point, even if they use a heterogeneous mix of AIMD, AIAD, MIMD, MIAD or other step functions.  Convergence behavior does depend on the choice of utility function, however.

\vspace{-4mm}
\subsection{Deployment}
\vspace{-2mm}

Despite being a significant shift in the congestion control architecture, PCC needs only isolated changes.  \textbf{No router support:} unlike ECN, XCP, and RCP, there are no new packet fields to be standardized and inspected, calculated upon, and modified by routers.  \textbf{No new protocol:} The packet format and semantics can simply remain as TCP (SACK, hand-shaking and etc.).  \textbf{No receiver change:}  TCP SACK is enough feedback. What PCC does change is the control algorithm within the sender, where the new intelligence lies.

The remaining concern is how PCC safely replaces and interacts with TCP. We observe that there are many scenarios where critical applications suffer severely from TCP's poor performance and PCC can be safely deployed by \emph{fully replacing} or \emph{being isolated from} TCP. First, \textbf{when a network resource is owned by a single entity} or can be reserved for it, the owner can replace TCP entirely with PCC. For example, some Content Delivery Network (CDN) providers use dedicated network infrastructure to move large amounts of data across continents~\cite{limelight, level3}, and scientific institutes can reserve bandwidth for exchanging huge scientific data globally~\cite{oscars}. Second, PCC can be used in challenging network conditions \textbf{where per-user or per-tenant resource isolation is enforced} by the network. Satellite Internet providers are known to use per-user bandwidth isolation to allocate the valuable bandwidth resource~\cite{tellitec}. For data centers with per-tenant resource isolation~\cite{splendidisolation, faircloud, elasticswitch}, an individual tenant can use PCC safely within its virtual network to address problems such as incast and improve data transfer performance between data centers.

The above applications, where PCC can fully replace or is isolated from TCP, are a significant opportunity for PCC. But in fact, \textbf{PCC does not depend on any kind of resource isolation to work.} It is possible that many individual users will, due to its significantly improved performance, decide to deploy PCC in the public Internet where unfriendly interaction with TCP is unavoidable. However, it turns out that PCC's unfriendliness to TCP is comparable to other selfish practices common today, so it is unlikely to make the ecosystem dramatically worse for TCP; see experiments in \S\ref{sec:unfriendly}.

\vspace{-6mm}
\subsection{Alternate utility functions}
\vspace{-2mm}
\label{sec:diffutility}
The above discussion assumes the ``safe'', general-purpose utility function of \S\ref{sec:safety}.
But a unique feature of PCC is that with flow-level fair queuing (FQ) in the network, applications can choose different utility functions to express heterogeneous optimization objectives (e.g. latency vs. throughput). Sivaraman et al.~\cite{nosilverbullet} recently observed that TCP has to rely on different in-network active queue management (AQM) mechanisms to cater to different applications' objectives (e.g. latency vs. throughput sensitivity) because \emph{even with FQ}, TCP is blind to applications' objectives. PCC opens a new level of flexibility with pluggable utility functions as we show in~\S~\ref{sec:eval-fq}, and thus can avoid the complexity and cost of programmable AQMs~\cite{nosilverbullet}.

\vspace{-4mm}
\section{Prototype Design}
\vspace{-3mm}
\label{sec:prototype}

\begin{figure}[t]
\label{fig:arch}
\centering
{{\includegraphics[width=2.5in]{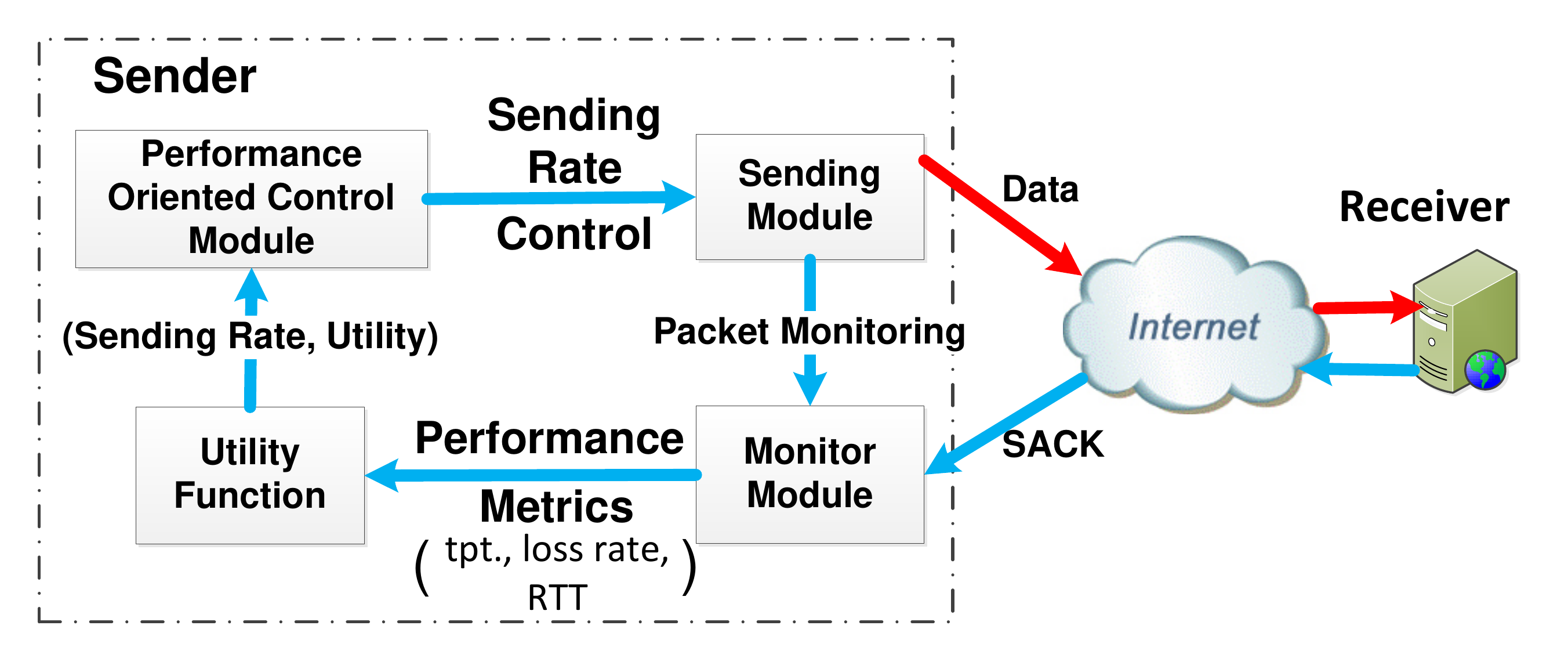}}
\vspace{-3mm}
\caption{\small \em \label{bbcc-prototype-arch}
PCC prototype design}
}
\vspace{-5mm}
\end{figure}

We implemented a prototype of PCC in user space on top of the TCP skeleton in the UDT~\cite{udt} package. Fig.~\ref{bbcc-prototype-arch} depicts our prototype's software components.

\vspace{-4mm}
\subsection{Performance Monitoring}
\label{sec:performancemonitoring}
\vspace{-3mm}

As described in ~\S\ref{sec:new_arch} and shown in Fig.~\ref{monitor}, the timeline is sliced into chunks of duration of $T_m$ called the \emph{Monitor Interval} (MI). When the Sending Module sends packets (new or retransmission) at a certain sending rate instructed by the Performance-oriented Rate Control Module, the Monitor Module will remember what packets are sent out during each MI. As the SACK comes back from receiver, the Monitor will know what happened (received? lost? RTT?) to each packet sent out during an MI. Taking the example of Fig.~\ref{monitor}, the Monitor knows what packets were sent during MI$1$, spanning $T_0$ to $T_0+T_m$, and at time $T_1$, approximately one RTT after $T_0+T_m$, it gets the SACKs for all packets sent out in MI$1$. The Monitor aggregates these individual SACKs to meaningful performance metrics including throughput, loss rate and average RTT. The performance metrics are then combined by a utility function and unless otherwise stated, we use the utility function of~\S\ref{sec:safety}. The end result of this is that we associate a control action of each MI (sending rate) with its performance result (utility). This information will be used by the performance oriented control module.


To ensure there are enough packets in one monitor interval, we set $T_m$ to the maximum of (a) the time to send 10 data packets and (b) a uniform-random time in the range $[1.7, 2.2]$ RTT. Again, we want to highlight that PCC \emph{does not} pause sending packets to wait for performance results, and it \emph{does not} decide on a rate and send for a long time; packet transfer and measurement-control cycles are truly continuous along each MI.

In some cases, the utility result of one MI can come back in the middle of another MI and the control module can decide to change sending rate after processing this result. As an optimization, PCC will immediately change the rate and ``re-align'' the current MI's starting time with the time of rate change without waiting for the next MI.


\begin{figure}[t]
\label{monitor}
\centering
{{\includegraphics[width=2.5in]{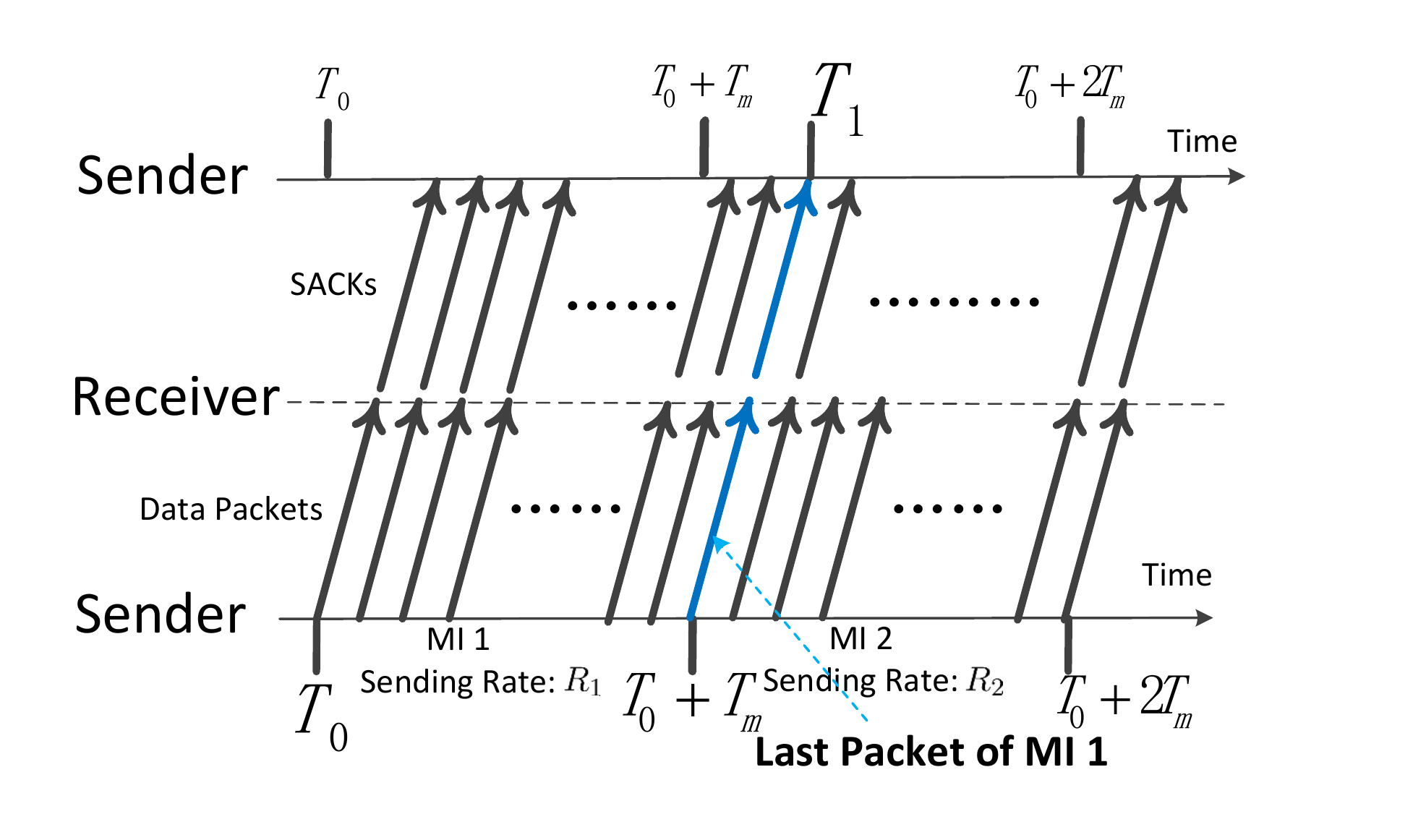}}
\vspace{-3mm}
\caption{\small \em \label{monitor}
Performance Monitoring Process}
}
\vspace{-5mm}
\end{figure}

\vspace{-4mm}
\subsection{Control Algorithm}
\vspace{-3mm}
\label{sec:controlalgorithm}
We propose a practical control algorithm with the gist of the simple control algorithm described in~\S\ref{sec:safety}.

\textbf{Starting State:} PCC starts at rate $2\cdot MSS/RTT$ and doubles its rate at each consecutive monitor interval (MI), like TCP. Unlike TCP, PCC does not exit this starting phase because of a packet loss. Instead, it monitors the utility result of each rate doubling action. Only when the utility decreases, PCC exits the starting state, returns to the previous rate which had higher utility (i.e., half of the rate), and enters the \emph{Decision Making State}. PCC could use other more aggressive startup strategies, but such tweaks could be applied to TCP as well.



\textbf{Decision Making State:} Assume \name is currently at rate $r$. To decide which direction and amount to change its rate, PCC conducts \textbf{multiple randomized controlled trials (RCTs)}. PCC takes four consecutive MIs and divides them into two paris (2 MIs each). For each pair, PCC attempts a slightly higher rate $r(1+\eps)$ and slightly lower rate $r(1-\eps)$, each for one MI, in random order. After the four consecutive trials, PCC changes the rate back to $r$ and keeps aggregating SACKs until the Monitor generates utility value for these four trials. For each pair $i \in {1,2}$, PCC gets two utility measurements $U_i^+, U_i^-$ corresponding to $r(1+\eps), r(1-\eps)$ respectively. If the higher rate consistently has higher utility ($U_i^+ > U_i^-~\forall i \in\{1,2\}$), then \name adjusts its sending rate to $r_{new}=r(1+\eps)$; and if the lower rate consistently has higher utility then \name picks $r_{new}=r(1-\eps)$. However, if the results are inconclusive, e.g. $U_1^+ > U_1^-$ but $U_2^+ < U_2^-$, PCC stays at its current rate $r$ and re-enters the Decision Making State with larger experiment granularity, $\eps = \eps + \eps_{min}$. The granularity starts from $\eps_{min}$ when it enters the decision making system for the first time and will increase up to $\eps_{max}$ if the process continues to be inconclusive. This increase of granularity helps PCC avoid getting stuck due to noise. Unless otherwise stated, we use $\eps_{min} = 0.01$ and $\eps_{max}=0.05$.

\textbf{Rate Adjusting State:} Assume the new rate after Decision Making is $r_0$ and $dir = \pm 1$ is the chosen moving direction. In each MI, \name adjusts its rate in that direction faster and faster, setting the new rate $r_n$ as: $r_{n} = r_{n-1}\cdot (1 + n \cdot \eps_{min} \cdot dir)$. However, if utility falls, i.e. $U(r_{n})<U(r_{n-1})$, PCC reverts its rate to $r_{n-1}$ and moves back to the \emph{Decision Making State}.


\vspace{-4mm}
\section{Evaluation}
\label{sec:eval}

\vspace{-2mm}

We demonstrate PCC's architectural advantages over the TCP family through diversified, large-scale and real-world experiments:
\S\ref{sec:eval-robust}: PCC achieves its design goal of \textbf{consistent high performance}.
\S\ref{sec:converge}: PCC can actually achieve much \textbf{better fairness and convergence/stability tradeoff} than TCP.
\S\ref{eval:deploy}: PCC is \textbf{practically deployable} in terms of flow completion time for short flows and TCP friendliness.
\S\ref{sec:eval-fq}: PCC has a huge potential to flexibly \textbf{optimize for applications' heterogenous objectives} with fair queuing in the network rather than more complicated AQMs~\cite{nosilverbullet}.

\vspace{-3mm}
\subsection{Consistent High Performance}
\label{sec:eval-robust}
\vspace{-3mm}

We evaluate PCC's performance under 8 real-world challenging network scenarios. With \emph{no algorithm tweaking for different scenarios and all experiments using the same ``safe'' utility function of \S\ref{sec:safety}}, in the first 7 scenarios, user-space PCC significantly outperforms in-kernel specially engineered TCP variants.


\vspace{-4mm}
\subsubsection{Big Data Transfer in the Wild}
\label{eval:planet}
\vspace{-2mm}

\begin{figure}[t]
\label{fig:planet}
\centering
{{\includegraphics[width=3in]{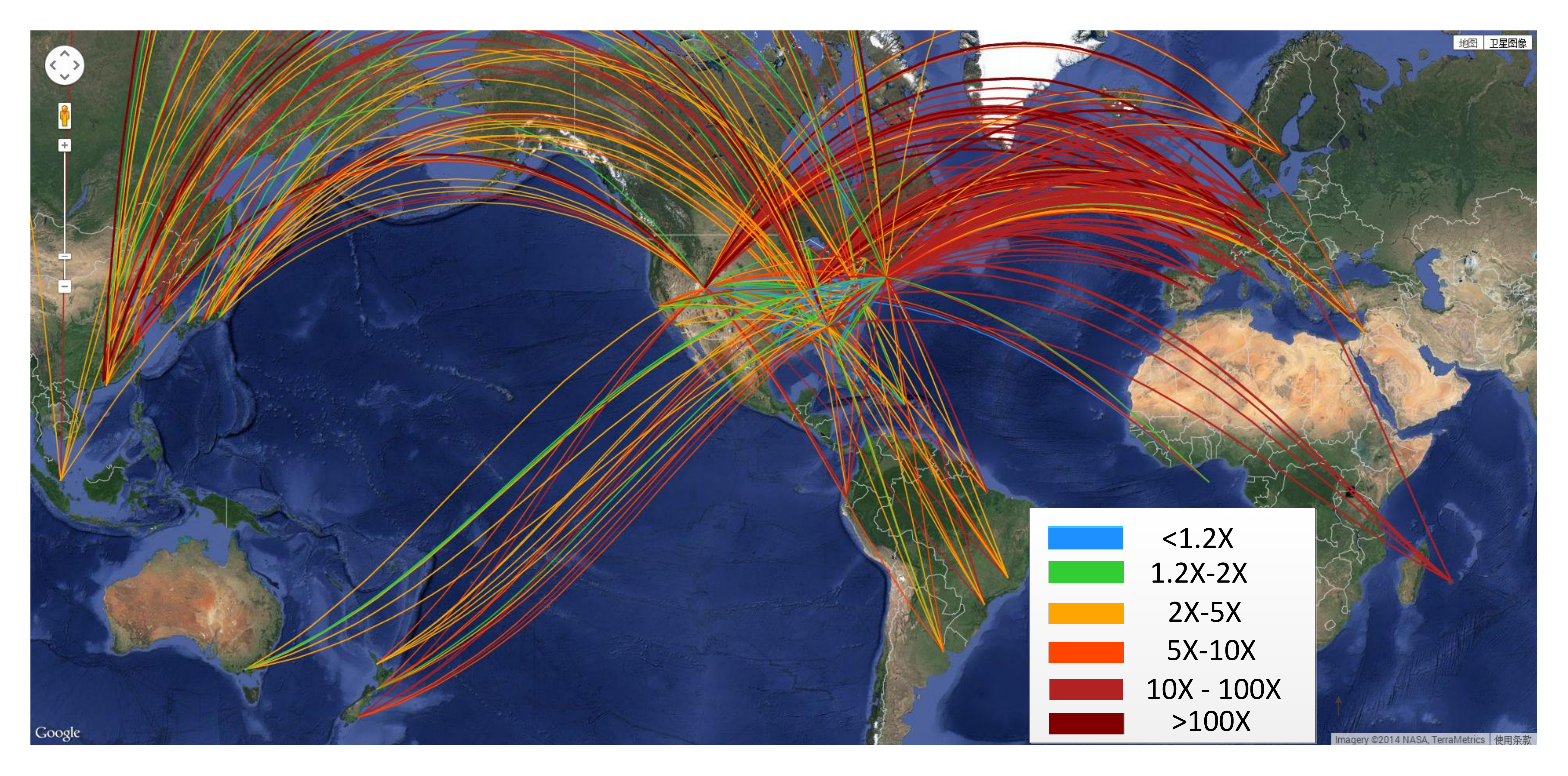}}
\vspace{-3mm}
\caption{\small \em \label{planet}
Large scale Internet experiment demonstrating PCC's performance improvement over TCP CUBIC}
}
\vspace{-5mm}
\end{figure}

Due to its complexity, the commercial Internet is the best place to test whether PCC can achieve consistently high performance. We deploy PCC's receiver on $85$ globally distributed PlanetLab~\cite{planetlab} nodes and senders on $6$ locations: five GENI~\cite{genitestbed} sites and our local server. These \textbf{$510$} sending-receiving pairs (Fig.~\ref{planet}) render a very diversified testing environment with BDP from $14.3$~KB to $18$~MB.

\begin{figure*}[t]
\hspace{-0.5cm}
  \begin{minipage}[t]{0.02\linewidth}~
  \end{minipage}
  \begin{minipage}[t]{0.3\linewidth}
    \centering
    \includegraphics[scale=.45]{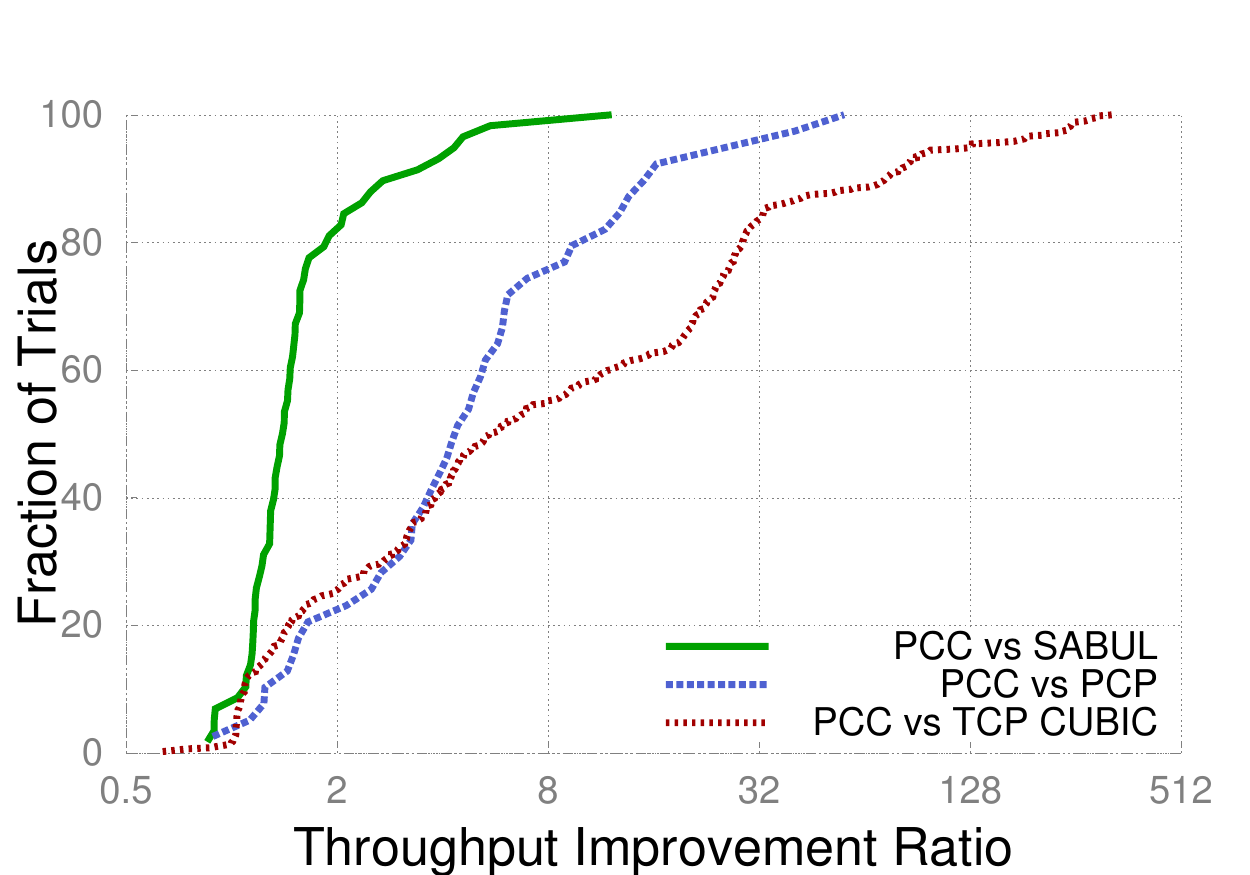}\\
    \vspace{-3mm}
    \caption{\small \em Across the public Internet, PCC has $\geq 10\times$ performance of CUBIC on $41\%$ of tested pairs}
    \label{cdf}
  \end{minipage}
    \begin{minipage}[t]{0.02\linewidth}~
  \end{minipage}
    \begin{minipage}[t]{0.3\linewidth}
    \centering
    \includegraphics[scale=.45]{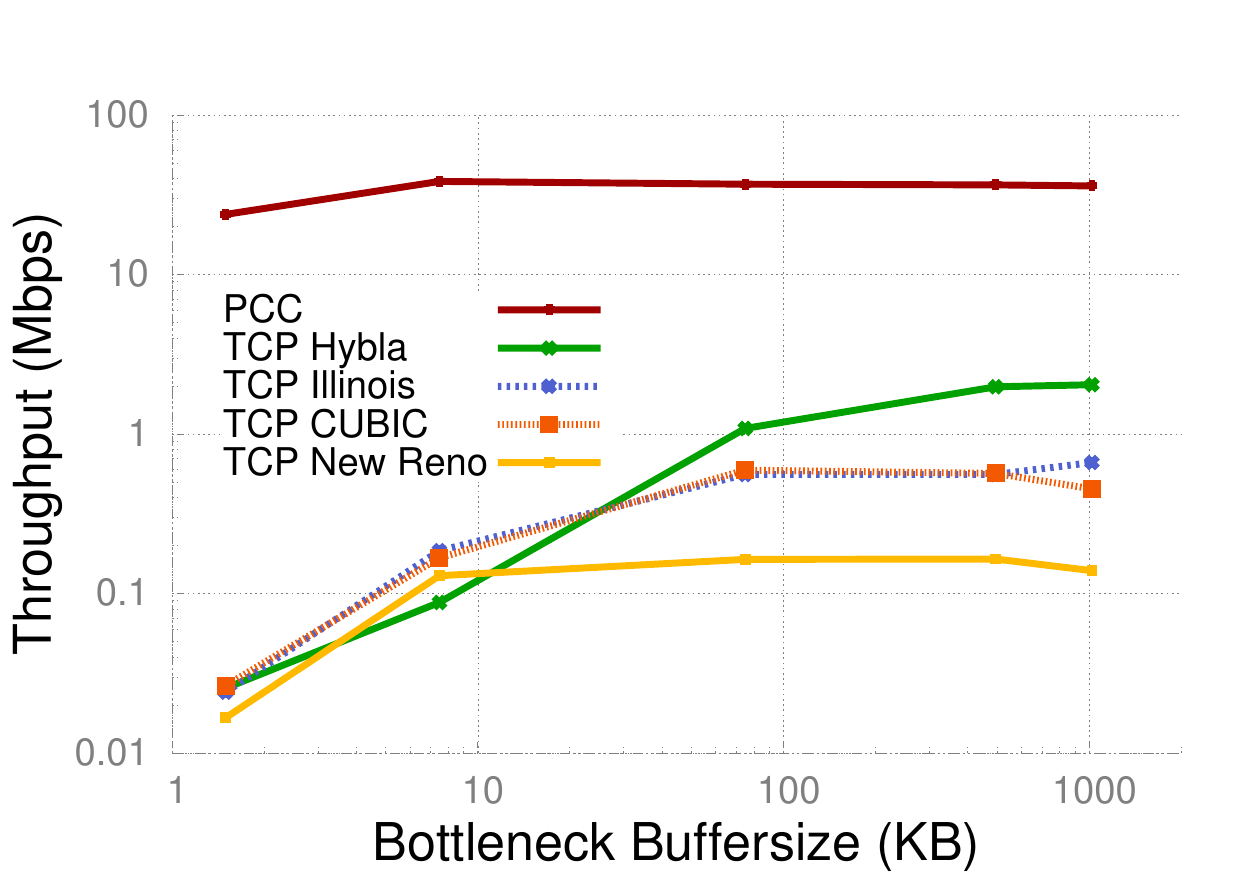}\\
    \vspace{-3mm}
    \caption{\small \em PCC outperforms special TCP modifications on emulated satellite links}
    \label{satellite}
  \end{minipage}
    \begin{minipage}[t]{0.02\linewidth}~
  \end{minipage}
  \begin{minipage}[t]{0.3\linewidth}
    \centering
    \includegraphics[scale=.45]{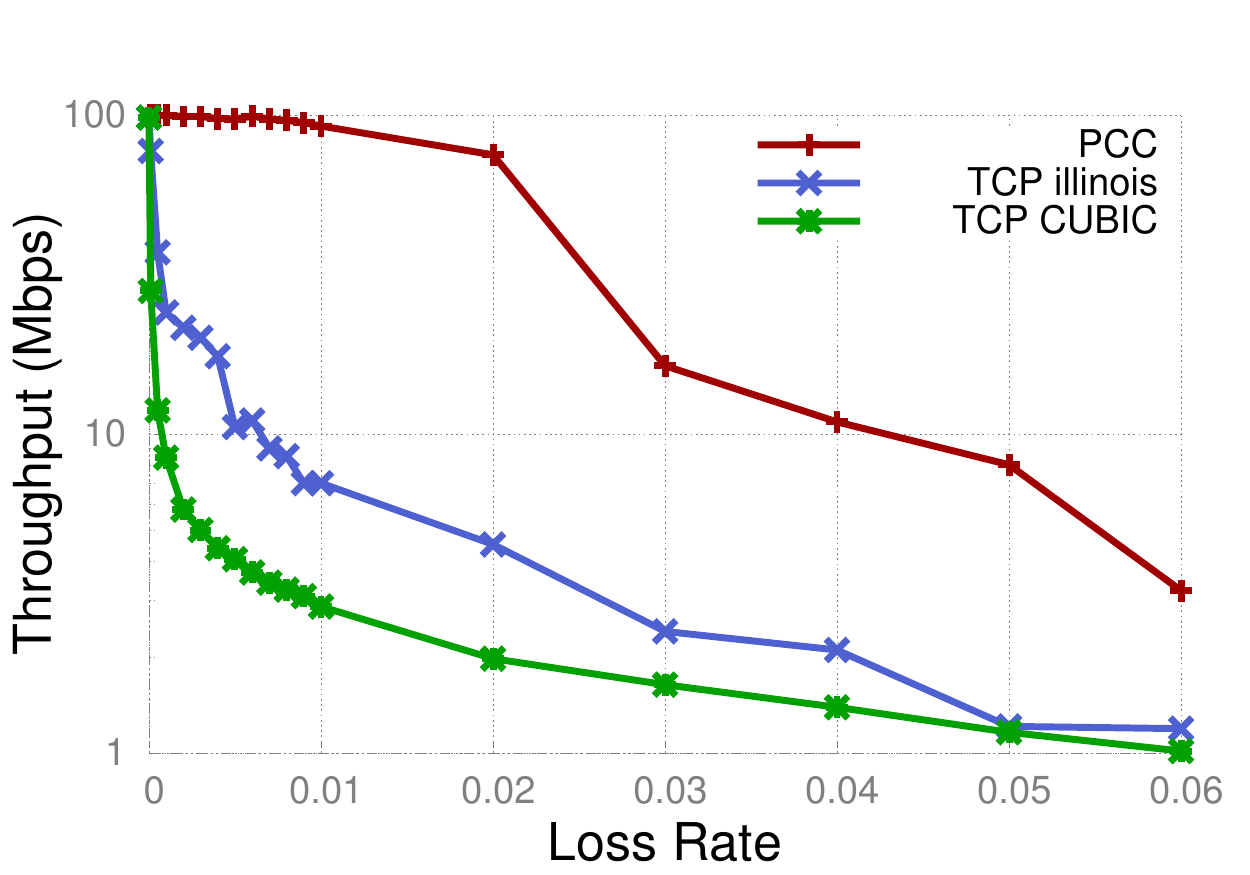}\\
    \vspace{-3mm}
    \caption{\small \em PCC is highly resilient to random loss compared to specially-engineered TCPs}
     \label{randomloss}
  \end{minipage}

\vspace{-3mm}
\end{figure*}

We first test PCC against TCP CUBIC, the Linux kernel default since 2.6.19; and also against a special modification of TCP for high BDP links. For each sender-receiver pair, we run TCP iperf between them for 100 seconds, wait for 500 seconds and then run PCC for 100 seconds to compare their average throughput.

PCC yields $5.52\times$ higher throughput than TCP CUBIC at the median (Fig.~\ref{cdf}). On $41\%$ of sender-receiver pairs, PCC's improvement is more than $10\times$; more than $50\times$ is not rare and it can be up to $300\times$. This is a conservative result because 4 GENI sites have 100Mbps bandwidth limits on their Internet uplinks. Thus, even CUBIC, a special modification for high BDP, fails to adapt to these complex real world networks and renders very poor performance.

We also tested two other non-TCP transport protocols on smaller scale experiments: the public releases of PCP~\cite{pcp, pcpimplementation} (43 sending receiving pairs) and SABUL~\cite{udt} (85 sending receiving pairs). PCP uses packet-trains~\cite{pathload} to probe available bandwidth. However, as discussed more in \S\ref{sec:rel}, this bandwidth probing is different from PCC's control based on empirically observed action-utility pairs, and contains unreliable assumptions that can yield very inaccurate sample results.  SABUL, widely used for scientific data transfer, packs a full set of boosting techniques: packet pacing, latency monitoring, random loss tolerance, etc. However, SABUL still mechanically hardwires control action to packet-level events.

Fig.~\ref{cdf} shows PCC outperforms PCP \footnote{$initial-rate=1Mbps$, $poll-interval =100 \mu s$. PCP in many cases abnormally slows down (e.g. 1 packet per 100ms). We have not determined whether this is an implementation bug in PCP or a more fundamental deficiency. To be conservative, we excluded all such results from the comparison.} by $4.58\times$ at the median and $15.03\times$ at the $90$th percentile, and outperforms SABUL by $1.41\times$ at the median and $3.39\times$ at the 90th percentile. SABUL shows an unstable control loop: it aggressively overshoots the network and than deeply falls back. On the other hand, PCC stably tracks the optimal rate. As a result, SABUL suffers from $11.5\%$ loss on average compared with PCC's $3.1\%$ loss.

We believe the major part of this performance gain is because PCC is able to make better decisions than the TCP-based hardwired-mapping protocols under real Internet conditions. We argue the gain is not simply due to PCC's TCP unfriendliness for the following reasons: \textbf{a.} As shown in \S\ref{sec:unfriendly}, PCC is more TCP-friendly than a bundle of 10 parallel TCP flows and $41\%$ of the experiments show larger than $10\times$ improvement; \textbf{b.} If PCC often gains more than $10\times$ performance by crowding out TCP flows, it means that most tested links are highly congested most of the time which conflicts with the general presumption that Internet links are mostly idle. \textbf{c.} PCC is less aggressive than SABUL, but still performs better. \textbf{d.} We conducted an additional experiment with 32 receiving nodes and only initiated a trial if PlanetLab monitors showed  $\leq 500$Kbps incoming traffic during the last one minute, to avoid possible congestion on the edge. The results show an improvement of $35\times$\footnote{Note that nodes with monitoring capability are in Europe and thus tend to have higher BDP.} at the median.



\vspace{-4mm}
\subsubsection{Inter-Data Center Environment}
\vspace{-2mm}

We next evaluate PCC's performance on inter-data center, cross-continent scientific data transfer~\cite{esnet} and dedicated CDN backbones~\cite{level3} where \textbf{network resource can be isolated or reserved for a single entity}.

The GENI testbed~\cite{genitestbed}, which has reservable bare-metal servers across the U.S. and reservable bandwidth~\cite{internet2ion} over the Internet2 backbone, provides us a representative environment for this evaluation. We choose \textbf{9 pairs} of GENI sites (with name shown in Table.~\ref{table:geni}) and \textbf{reserve 800Mbps end-to-end dedicated bandwidth} between each pair. We compare PCC, SABUL, TCP CUBIC and TCP Illinois over 100-second runs.

As shown in Table~\ref{table:geni}, PCC significantly outperforms TCP Illinois, by $5.2\times$ on average and up to $7.5\times$. It is surprising that even in this very clean network, specially optimized TCPs still perform far from optimal. We believe some part of the gain is because the bandwidth-reserving rate limiter has a small buffer and TCP will overflow it, unnecessarily decreasing rate. On the other hand, PCC continuously track the optimal sending rate by continuously measuring the performance. TCP pacing will not resolve this problem as shown in \S\ref{eval:shallowbuffer}.

\begin{table} [hbp] \scriptsize
  \centering
  \caption{PCC significantly outperforms TCP in inter-data center environments. RTT is in msec; throughput in Mbps.}
    \begin{tabular}{rrrrrrr}
    \toprule
    Transmission Pair & RTT & PCC & SABUL & CUBIC & Illinois \\
    \midrule
    GPO $\to$ NYSERNet & 12.1 & 818  & 563  & 129  & 326\\
    GPO $\to$ Missouri & 46.5 & 624 & 531 & 80.7 & 90.1\\
    GPO $\to$ Illinois & 35.4 & 766 & 664  & 84.5  & 102\\
    NYSERNet $\to$ Missouri & 47.4 & 816 & 662 & 108 & 109\\
    Wisconsin $\to$ Illinois & 9.01 & 801 & 700  & 547   & 562\\
    GPO $\to$ Wisc. & 38.0 & 783  & 487  & 79.3  & 120\\
    NYSERNet $\to$ Wisc. & 38.3 & 791 & 673 & 134 & 134\\
    Missouri $\to$ Wisc. & 20.9 & 807 & 698  & 259  & 262\\
    NYSERNet $\to$ Illinois & 36.1 & 808 & 674& 141 & 141\\
    \bottomrule
    \end{tabular}%
  \label{table:geni}%
  \vspace{-5mm}
\end{table}%

\vspace{-4mm}
\subsubsection{Satellite Links}
\vspace{-2mm}

Satellite Internet is widely used for critical missions such as emergency and military communication and Internet access for rural areas. Because TCP suffers from severely degraded performance on satellite links that have excessive latency (600ms to 1500ms RTT~\cite{vsatlatency}) and relatively high random loss rate~\cite{winds}, special modifications of TCP (Hybla, Illinois) were proposed and even special infrastructure has been built~\cite{vsat, satellitetutorial}.

We test PCC against TCP Hybla~(widely used in real-world satellite communication), Illinois and CUBIC under emulated satellite links on Emulab parameterized with the real-world measurement for the WINDs satellite Internet system~\cite{winds}. The satellite link has \textbf{$800ms$ RTT, $42$Mbps capacity and $0.74\%$ random loss}. As shown in Fig.~\ref{satellite}, we vary the bottleneck buffer from $1.5$KB to $1$MB and compare PCC's average throughput against different TCP variants with $100$ second trials. PCC achieves $90\%$ of optimal throughput even with only $7.5KB$ buffer (5 packets) at the bottleneck. However, even with $1MB$ buffer, the widely used TCP Hybla can only achieve $2.03$Mbps which is $17\times$ worse than PCC. TCP Illinois, which is designed for high random loss tolerance, performs $54\times$ worse than PCC with $1MB$ buffer.

\vspace{-4mm}
\subsubsection{Unreliable Lossy Links}
\vspace{-2mm}

Lossy links in today's network are not uncommon: wireless links are often unreliable and very long wired network paths can also have random loss caused by unreliable infrastructure\footnote{according to our industry contact involving cross continent data delivery}. To further quantify the effect of random loss, we set up a link on Emulab with \textbf{100Mbps bandwidth, 30ms RTT and varying loss rate} on both forward and backward paths. As shown in Fig.~\ref{randomloss}, PCC can reach $> 95\%$ of achievable throughput capacity until loss rate reaches $1\%$ and shows relatively graceful performance degradation from $95\%$ to $74\%$ of capacity as loss rate increases to  $2\%$. However, TCP's performance collapses very quickly: CUBIC's performance collapse to $10\times$ smaller than PCC with only $0.1\%$ loss rate and $37\times$ smaller than PCC with $2\%$ random loss. TCP Illinois shows better resilience than CUBIC but throughput still degrades severely to less than $10\%$ of PCC's throughput with only $0.7\%$ loss rate and $16\times$ smaller than PCC with $2\%$ random loss. Again, PCC can endure random loss because it looks at real utility: unless link capacity is reached, a higher rate will always result in similar loss rate and higher throughput, which translates to higher utility.

PCC's performance does decrease to $3\%$ of the optimal achievable throughput when loss rate increases to $6\%$ because we are using the ``safe'' utility function of \S\ref{sec:safety} that caps the loss rate to $5\%$. We evaluate a utility function that endures excessive random loss later (\S\ref{eval:excessiveloss}).

\vspace{-4mm}
\subsubsection{Mitigating RTT Unfairness}
\vspace{-2mm}

\begin{figure*}[t]
\hspace{-0.5cm}
  \begin{minipage}[t]{0.02\linewidth}~
  \end{minipage}
  \begin{minipage}[t]{0.3\linewidth}
    \centering
    \includegraphics[scale=.45]{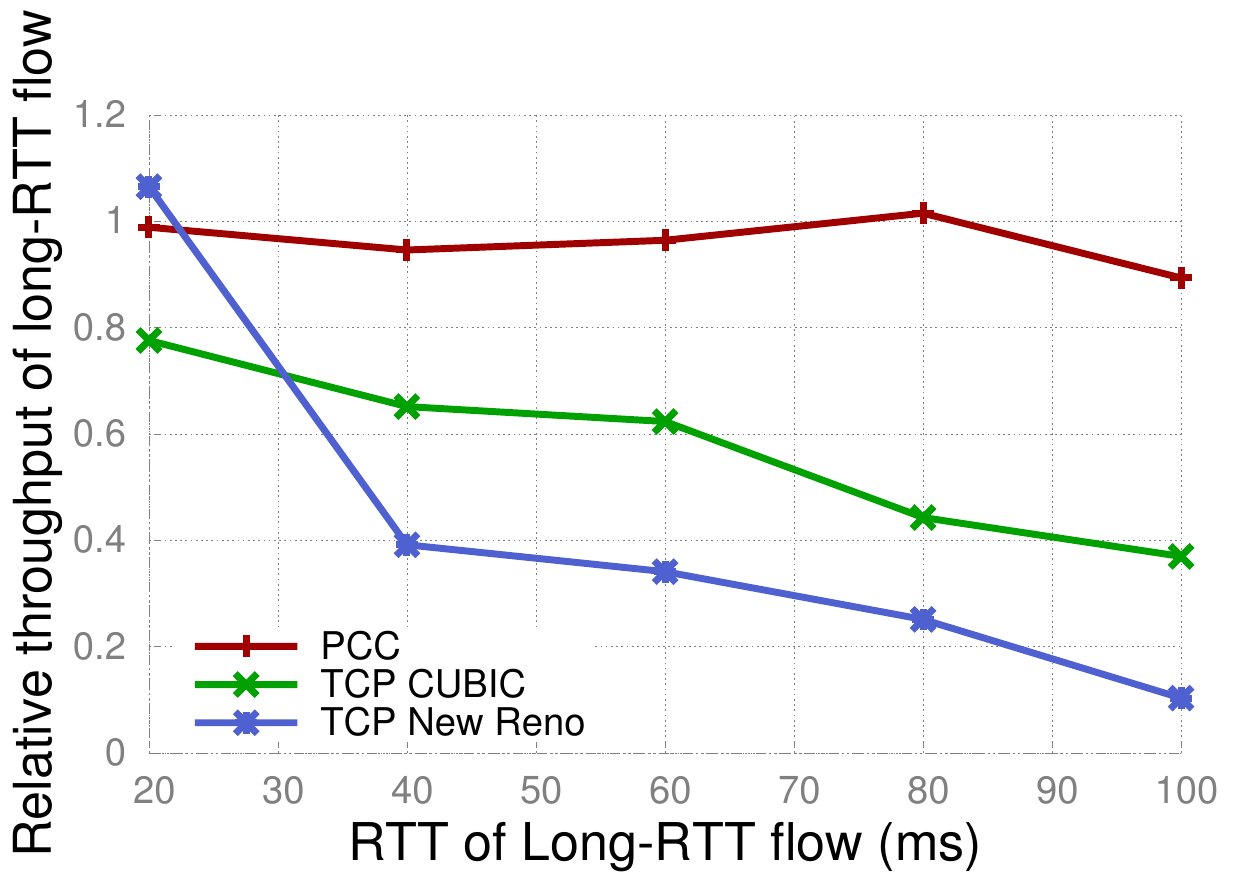}\\
    \vspace{-3mm}
    \caption{\small \em PCC achieves better RTT fairness comparing to specially engineered TCP}
    \label{unfairrtt}

  \end{minipage}
    \begin{minipage}[t]{0.02\linewidth}~
  \end{minipage}
    \begin{minipage}[t]{0.3\linewidth}
    \centering
\includegraphics[scale = .45]{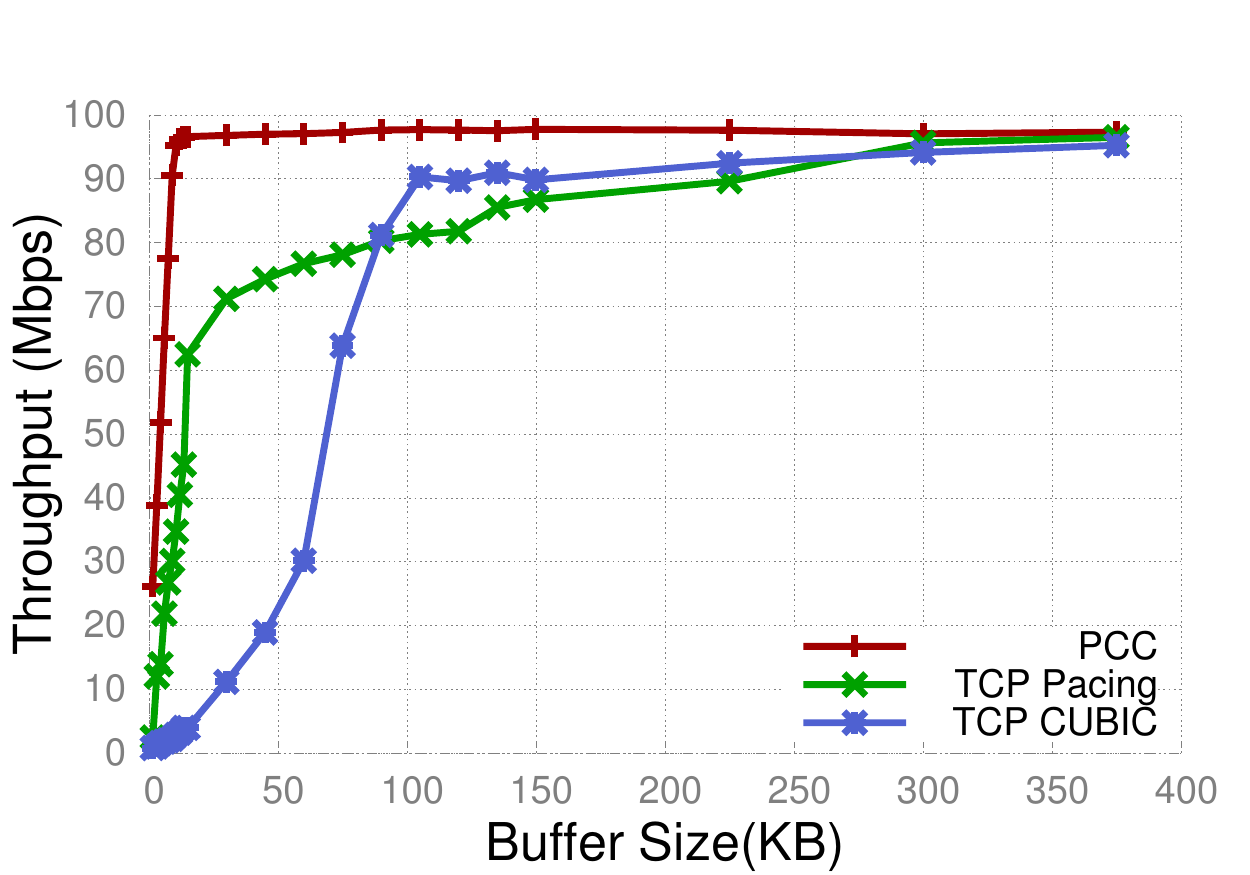}\\
\vspace{-3mm}
\caption{\small \em PCC efficiently utilizes shallow-buffered network}
\label{shallowbuffer}
  \end{minipage}
    \begin{minipage}[t]{0.02\linewidth}~
  \end{minipage}
  \begin{minipage}[t]{0.3\linewidth}
    \centering
    \includegraphics[scale=.45]{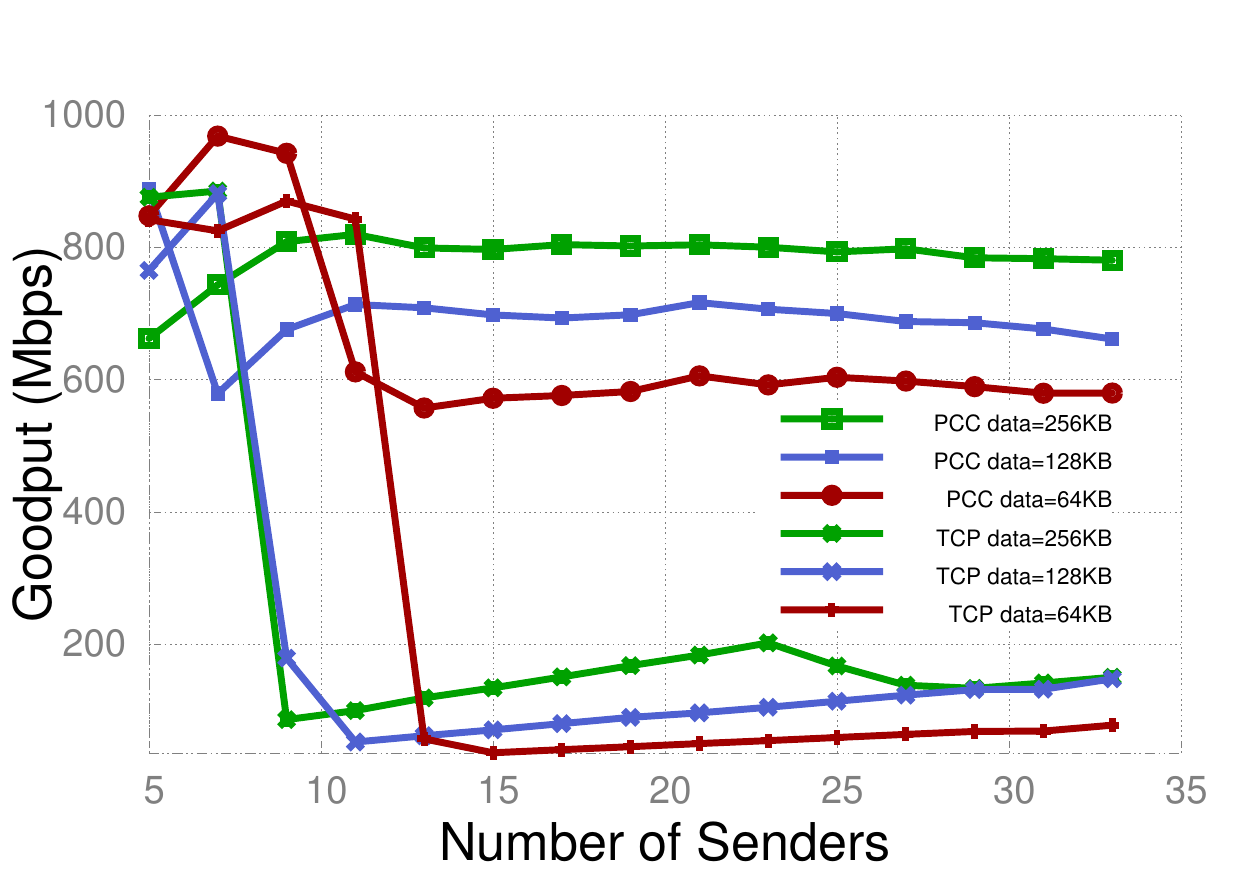}\\
    \vspace{-3mm}
    \caption{\small \em PCC largely mitigate TCP incast in data center}
    \label{goodput}

  \end{minipage}
\vspace{-5mm}
\end{figure*}

For unmodified TCP, short-RTT flows dominate long-RTT flows on throughput. Subsequent modifications of TCP such as CUBIC or Hybla try to mitigate this problem by making the expansion of the congestion window independent of RTT. However, the modifications cause new problems like parameter tuning (Hybla) and severely affect stability on high RTT links (CUBIC)~\cite{CUBIC}. Because PCC's convergence is based on real performance not the control cycle length, it acts as an architectural cure for the RTT unfairness problem. To demonstrate that, on Emulab we set \textbf{one short-RTT (10ms) and one long-RTT (varying from 20ms to 100ms) network path sharing the same bottleneck link of 100Mbit/s bandwidth} and buffer equal to the BDP of the short-RTT flow. We run the long-RTT flow first for 5s, letting it grab the bandwidth, and then let the short-RTT flow join to compete with the long-RTT flow for 500s and calculate the ratio of the two flows' throughput. As shown in Fig.~\ref{unfairrtt}, PCC achieves much better RTT fairness than New Reno and even CUBIC cannot perform as well as PCC.

\vspace{-4mm}
\subsubsection{Small Buffer on the Bottleneck Link}
\vspace{-2mm}
\label{eval:shallowbuffer}

TCP cannot distinguish between loss due to congestion and loss simply due to buffer overflow. In face of high BDP links, a shallow buffered router will keep chopping TCP's window in half and the recovery process is very slow. However, very large network buffers are undesirable due to increased latency. This conflict makes choosing the right buffer size for routers a challenging multi-dimensional optimization problem~\cite{appenzeller2004sizing, prasad2007router} for network operators to balance between throughput, latency, cost of buffer memory, degree of multiplexing, etc. The common practice of over-buffering networks, in the fear that an under-buffered router will drop packets or leave the network severely under-utilized, can result in a bufferbloat epidemic~\cite{bufferbloat}.

The complication of choosing the right buffer size would be much less painful if the transport protocol could efficiently utilize a network with very shallow buffers. Therefore, we test how PCC performs with  a tiny buffer and compare with TCP CUBIC, which is known to mitigate this problem. Moreover, to address the concern that the performance gain of PCC is merely due to PCC's use of packet pacing, we also test an implementation of TCP New Reno with pacing rate of $(congestion window)/(RTT)$. We set up on Emulab a network path with \textbf{30ms RTT, 100Mbps bottleneck bandwidth and vary the network buffer size from 1.5KB (one packet) to 375KB ($1 \times BDP$)} and compare the protocols' average throughput on this path over 100s.

As shown in~\ref{shallowbuffer}, PCC only requires $6\cdot MSS$ buffer to reach $90\%$ capacity. With the same buffer, CUBIC can only reach $2\%$ capacity and even TCP with packet pacing can only reach $30\%$. CUBIC requires $13\times$ more buffer than PCC to reach $90\%$ throughput and takes $36\times$ more buffer to close the $10\%$ gap. Even with pacing, TCP still requires $25\times$ more buffer than PCC to reach 90\% throughput. It is also critical to notice that PCC's throughput can reach $25\%$ of capacity with just a \emph{single-packet} buffer, $35\times$ higher throughput than TCP under the same scenario. The reason is that PCC constantly monitors the real achieved performance and steadily tracks its rate at the bottleneck rate without swinging up and down like TCP. That means with PCC, network operators can use shallow buffered routers to \textbf{get low latency without harming throughput.}

\vspace{-4mm}
\subsubsection{Rapidly Changing Networks}
\vspace{-2mm}

\begin{figure}[t]
\label{fig:dynamicnetwork}
\centering
{{\includegraphics[width=3.2in]{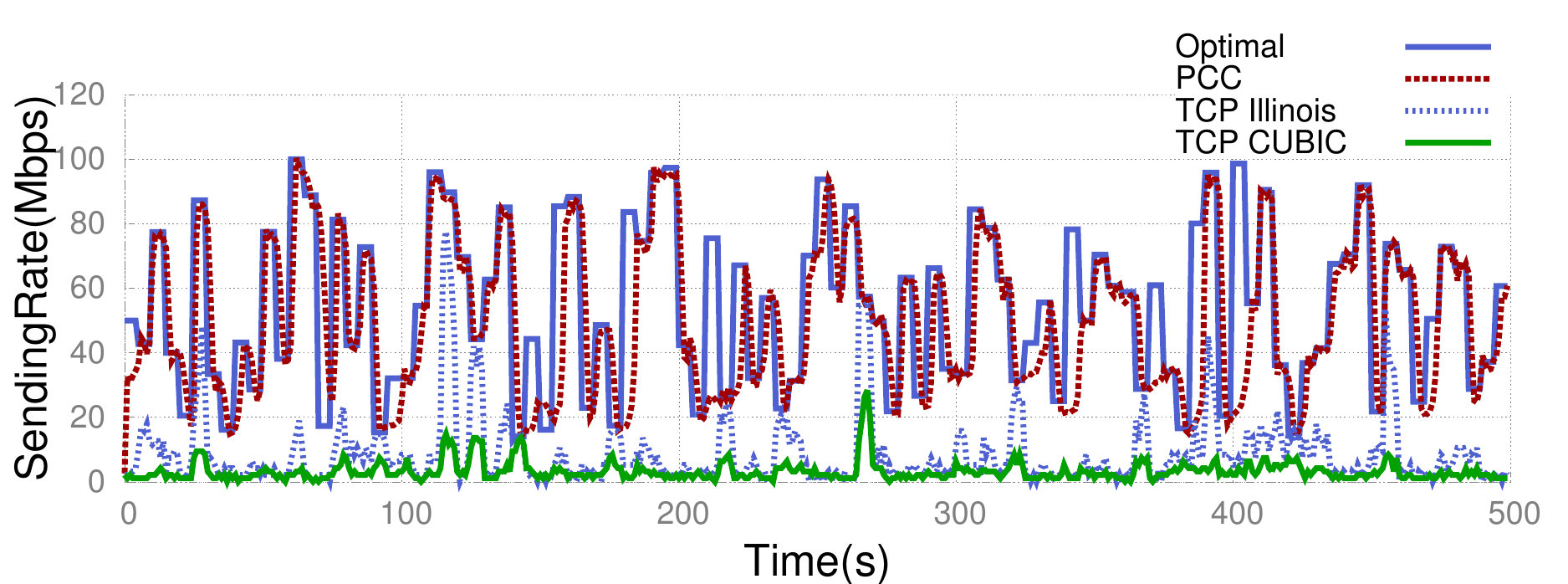}}
\caption{\small \em PCC can always track optimal sending rate even with drastically changing network conditions\label{dynamicnetwork}
}
}
\vspace{-5mm}
\end{figure}

The above mentioned evaluated scenarios are all ``static'' in the sense that network condition does not change dramatically during the test. Next, we study a rapidly changing network. We set up on Emulab a network path where  \textbf{available bandwidth, loss rate and RTT, are all changing every $5$ seconds.} Each parameter is chosen independently from a uniform random distribution with bandwidth ranging from $10$Mbps to $100$Mbps, latency from $10$ms to $100$ms and loss rate from $0\%$ to $1\%$.

Figure~\ref{dynamicnetwork} shows available bandwidth (optimal sending rate), and the sending rate of PCC, CUBIC and Illinois. Note that we show the PCC's control algorithm's decided sending rate (not its throughput) to get insight into how PCC handles network dynamics. We can see with all network parameters rapidly changing, PCC tracks the available bandwidth very well. On the other hand, the TCPs fail to handle this scenario. Over the course of the experiment (500s), PCC's throughput is $44.9$Mbps achieving $83\%$ of the optimal, while TCP CUBIC and TCP Illinois are $14\times$ and $5.6\times$ worse than PCC respectively.

\vspace{-4mm}
\subsubsection{Incast}
\vspace{-2mm}
\label{sec:incast}

Moving from wide-area networks to the data center, we now investigate TCP incast~\cite{understandingincast}, which occurs in high bandwidth and low latency networks when multiple senders send data to one receiver concurrently, causing throughput collapse. To solve the TCP incast problem, many protocols have been proposed, including ICTCP~\cite{ictcp} and DCTCP~\cite{dctcp}. Here, we demonstrate PCC can achieve high performance under incast without special-purpose algorithms. We deployed PCC on Emulab~\cite{emulab} with \textbf{33 senders and 1 receiver}.


Fig.~\ref{goodput} shows the goodput of PCC and TCP across various flow sizes and numbers of senders. Each point is the average of 15 trials.  When incast congestion begins to happen with roughly $\geq 10$ senders, PCC achieves roughly $60$-$80$\% of the maximum possible goodput, or $7$-$8\times$ that of TCP.  Note that ICTCP~\cite{ictcp} also achieved roughly $60$-$80$\% goodput in a similar environment. Also, DCTCP's goodput degraded with increasing number of senders~\cite{dctcp}, while PCC's is very stable.

\vspace{-3mm}
\subsection{Dynamic Behavior of Competing Flows}
\label{sec:converge}
\vspace{-2mm}

\begin{figure}[t]
\centering
\subfigure[PCC converges very stable]{
\includegraphics[width=3in]{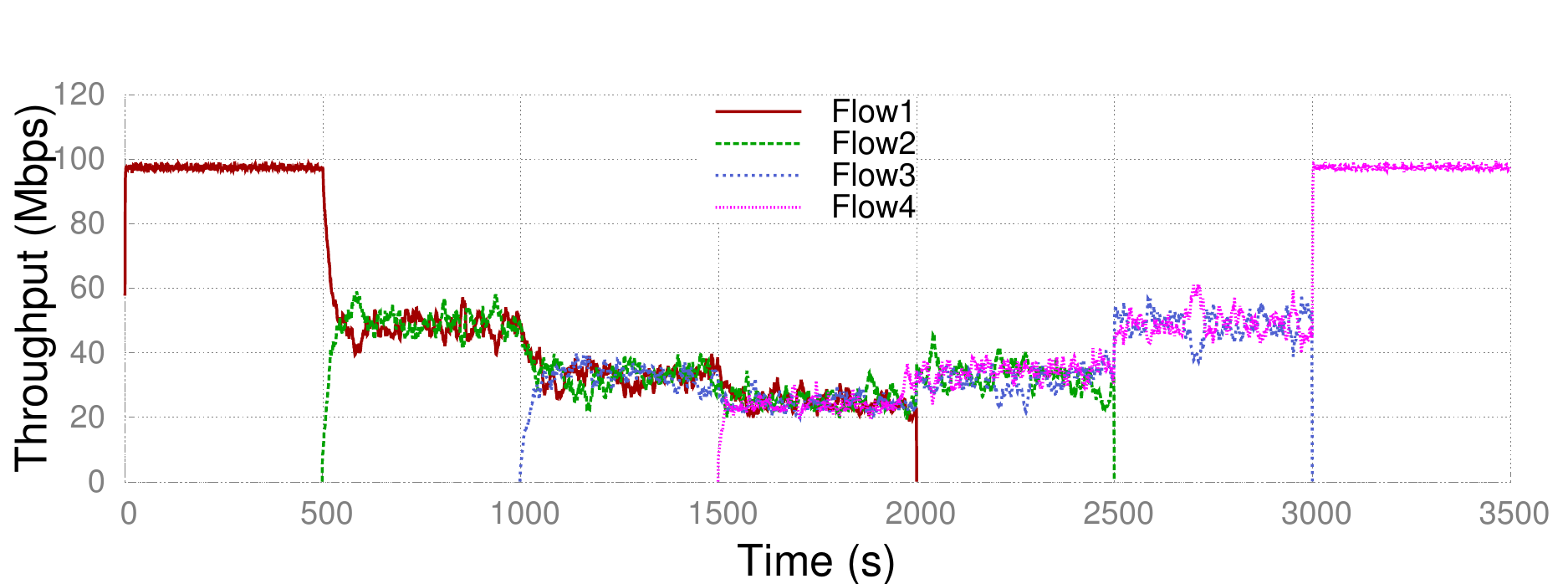}
}\\
\vspace{-6mm}
\subfigure[TCP CUBIC shows high rate variance and unfairness at short time scale]{
\includegraphics[width=3 in]{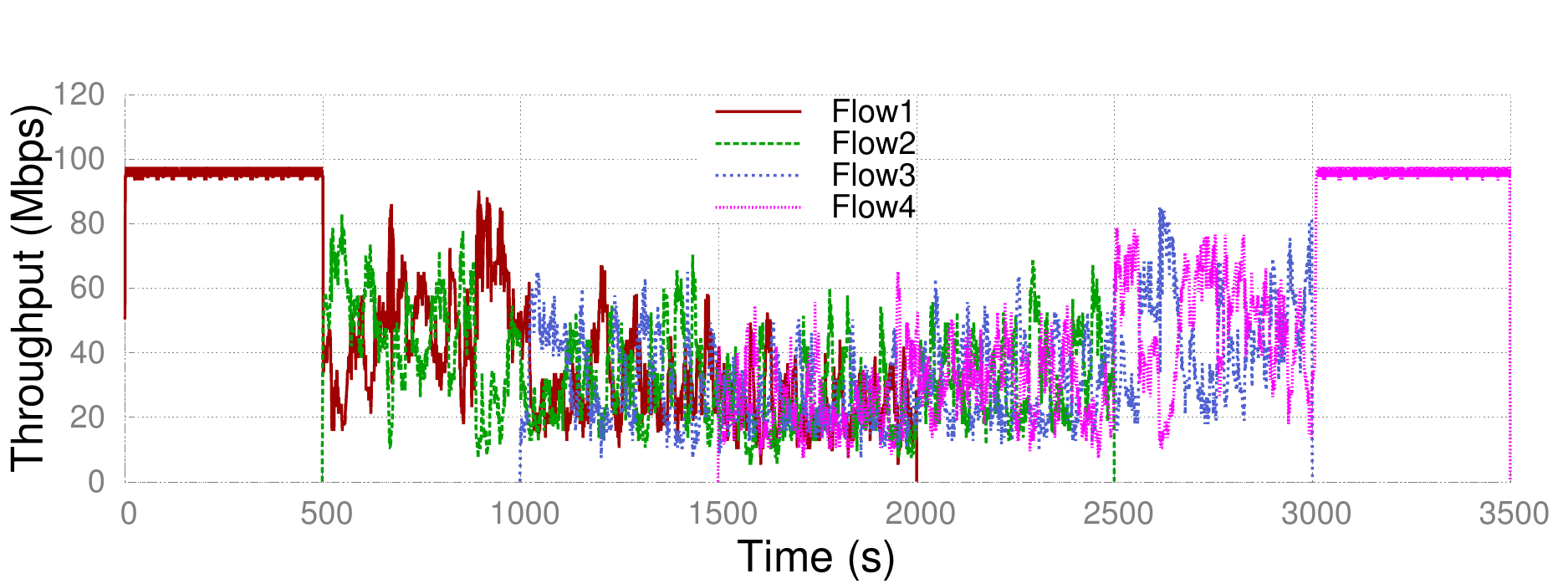}
}
\vspace{-4mm}
\caption{\small \em PCC converges much more stable than TCP CUBIC under FIFO queue}
\label{convergebbccfifo}
\vspace{-7mm}
\end{figure}

We proved in \S\ref{sec:safety} that with our ``safe'' utility function, competing PCC flows converge to a fair equilibrium from any initial state. In this section, we experimentally show that \textbf{PCC is much more stable, more fair and achieves a better tradeoff between stability and reactiveness than TCP.} PCC's stability can immediately translate to benefits for applications such as video streaming where stable rate in presence of congestion is desired~\cite{jiang2012improving}.
\vspace{-4mm}
\subsubsection{PCC is More Fair and Stable Than TCP}
\vspace{-2mm}
\label{sec:betterfair}

\begin{figure*}[t]
\hspace{-0.5cm}
  \begin{minipage}[t]{0.02\linewidth}~
  \end{minipage}
    \begin{minipage}[t]{0.3\linewidth}
    \centering
\includegraphics[scale = .45]{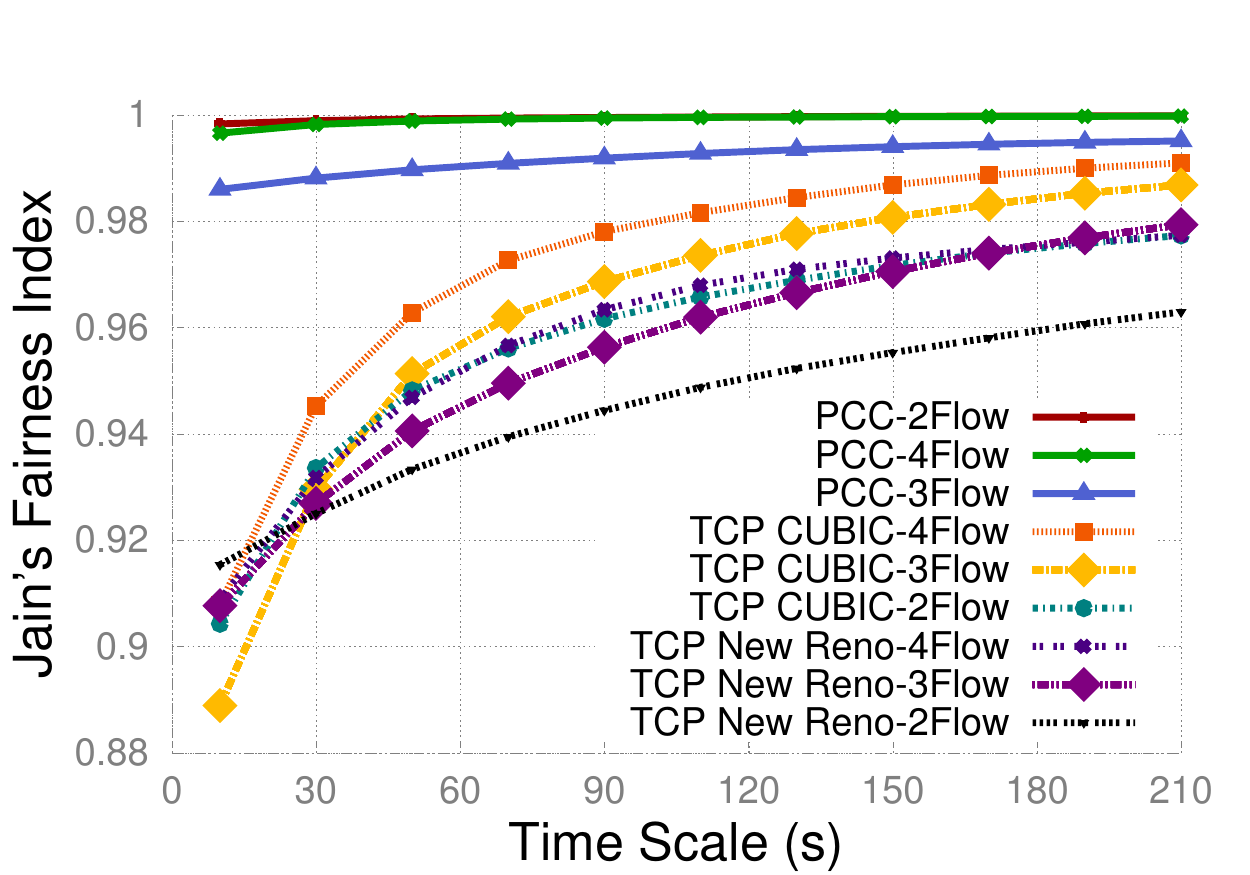}\\
\vspace{-3mm}
\caption{\small \em \label{fairness} PCC achieves better fairness in convergence than TCP CUBIC}
  \end{minipage}
    \begin{minipage}[t]{0.02\linewidth}~
  \end{minipage}
  \begin{minipage}[t]{0.3\linewidth}
    \centering
    \includegraphics[scale=.45]{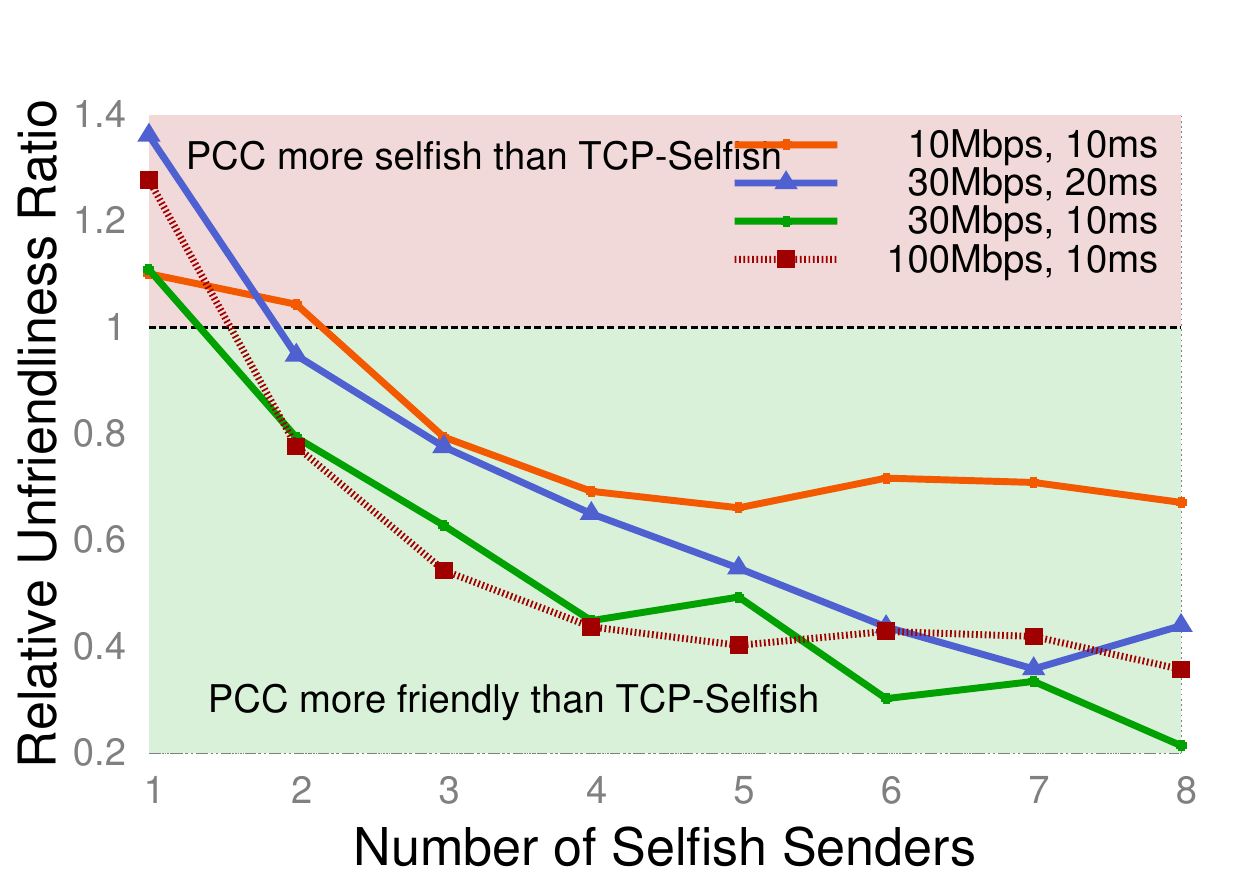}\\
    \vspace{-3mm}
    \caption{\small \em PCC's TCP unfriendliness is similar to common selfish practice}
    \label{unfriendliness}
  \end{minipage}
      \begin{minipage}[t]{0.02\linewidth}~
  \end{minipage}
    \begin{minipage}[t]{0.3\linewidth}
    \centering
\includegraphics[scale = .45]{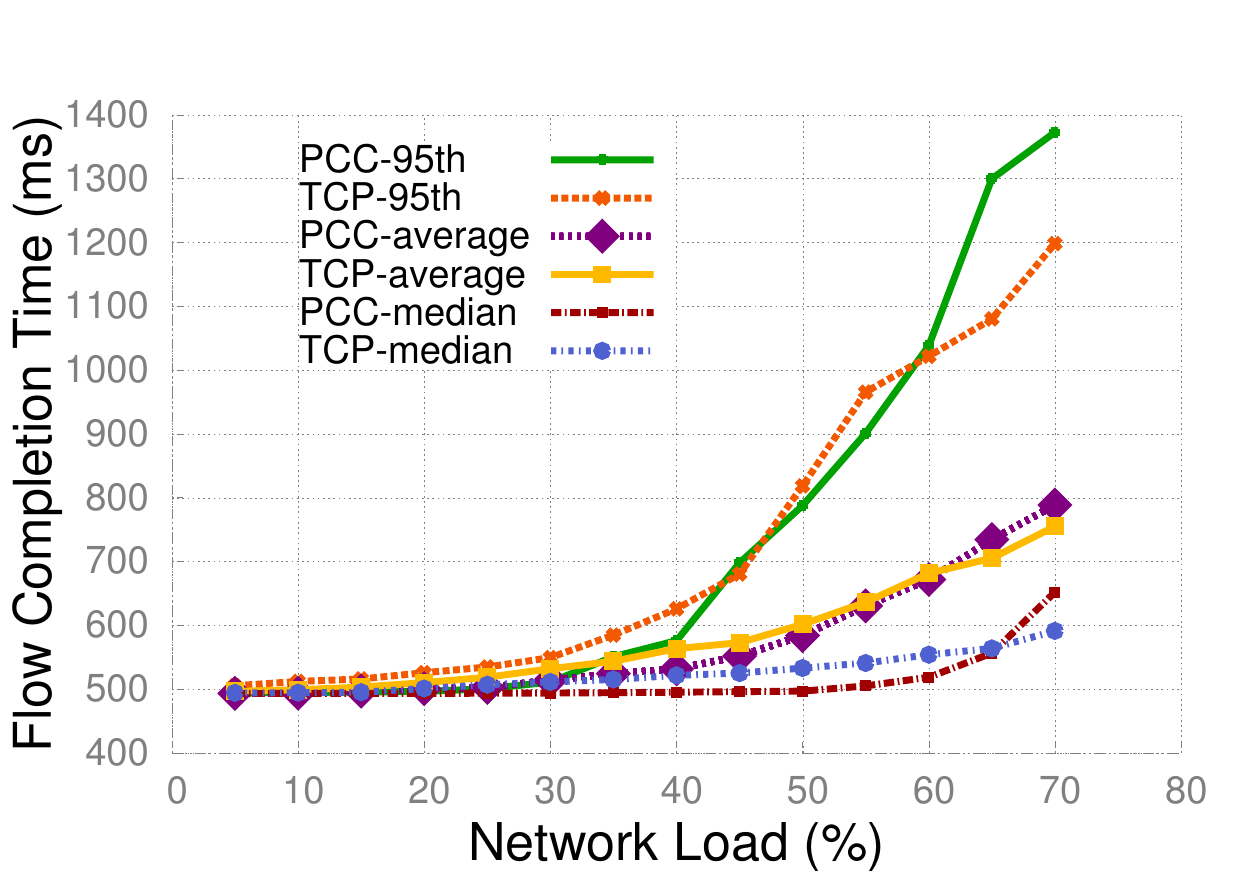}\\
\vspace{-3mm}
\caption{\small \em \label{shortflowfct}
PCC can achieve similar flow completion time for short flows comparing to TCP}
  \end{minipage}
\vspace{-5mm}
\end{figure*}

To compare PCC and TCP's convergence process in action, we set up a dumbbell topology on Emulab with \textbf{four senders and four receivers sharing a bottleneck link with $30$ms RTT, $100$Mbps bandwidth}. Bottleneck router buffer size is set to the BDP to allow CUBIC to reach full throughput.

The data transmission of the four pairs initiates sequentially with a $500$s interval and each pair transmits continuously for $2000$s. Fig.~\ref{convergebbccfifo} shows the rate convergence process for PCC and CUBIC respectively with $1$s granularity. It is visually obvious that PCC flows converge much more stably than TCP, which has surprisingly high rate variance. Quantitatively, we compare PCC's and TCP's fairness ratio (Jain's index) at different time scales (Fig.~\ref{fairness}). Selfishly competing PCC flows achieve better fairness than TCP at all time scales.

\vspace{-4mm}
\subsubsection{PCC has better Stability-Reactiveness tradeoff than TCP}
\vspace{-2mm}

\label{sec:tradeoff}

\begin{figure}[t]
\label{fig:randomloss}
\centering
{{\includegraphics[width=3.0in]{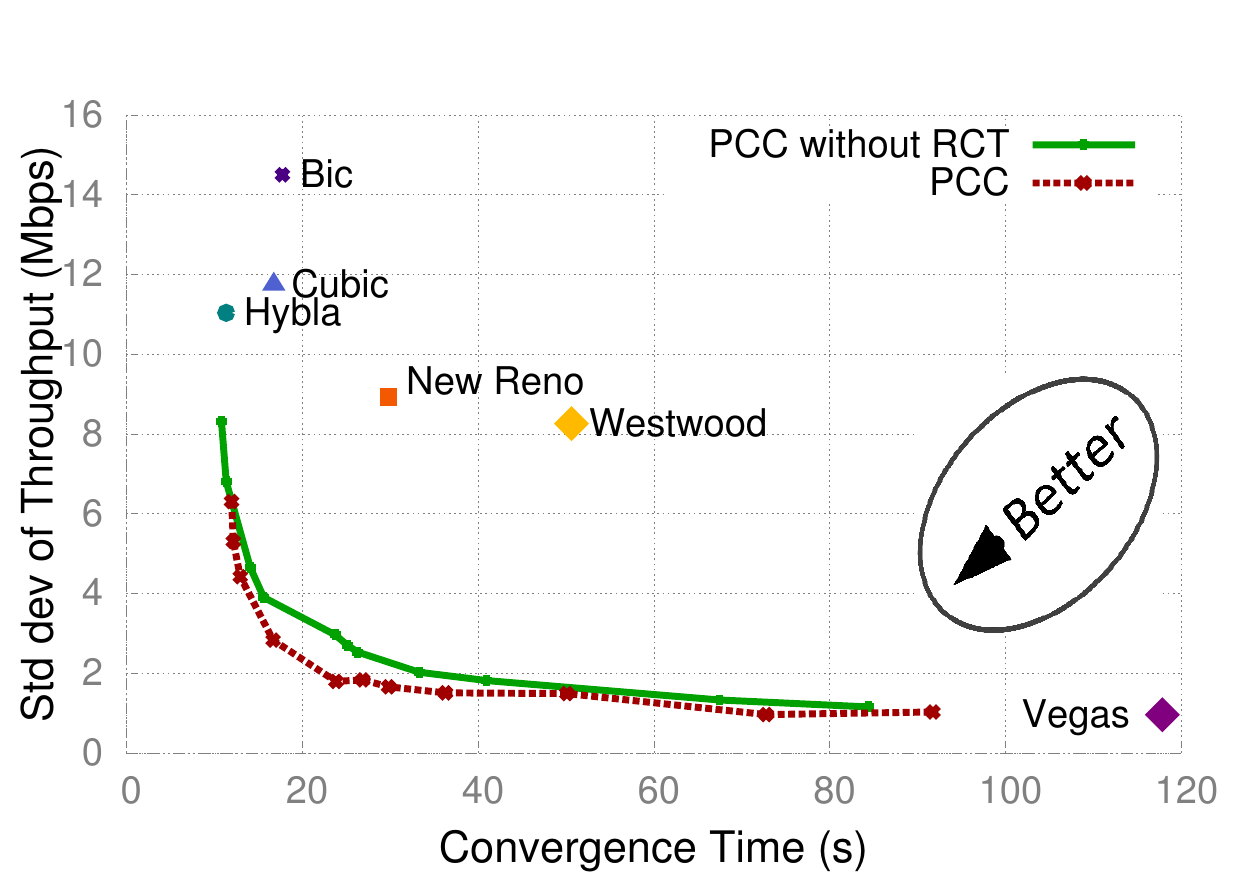}}
\caption{\small \em \label{tradeoff}
PCC has better reactiveness-stability tradeoff than TCP and RCT's benefit}
}
\vspace{-5mm}
\end{figure}

Intuitively, PCC's control cycle is ``longer'' than TCP due to performance monitoring. Is PCC's significantly better stability and fairness achieved by severely sacrificing convergence time?


We set up two sending-receiving pairs sharing a bottleneck link of 100Mbps and 30ms RTT. We conduct the experiment by letting the first flow, flow A, come in the network for 20s and then let the second flow, flow B, initiate. We define the convergence time in a ``forward-looking'' way: we say flow B's \emph{convergence time} is the smallest $t$ for which throughput in each second from $t$ to $t+5s$ is within $\pm 25\%$ of the ideal equal rate. We measure stability by measuring the standard deviation of throughput of flow B for $60s$ after convergence time.  All results are averaged over $15$ trials. PCC can achieve various points in the stability-reactiveness trade-off space by adjusting its parameter choice: higher step size $\eps_{min}$ and lower monitor interval $T_m$ result in faster convergence but higher throughput variance.  In Fig.~\ref{tradeoff}, we plot a trade-off curve for PCC by choosing a range of different settings of these parameters.\footnote{We first fix $\eps_{min}$ at $0.01$ and varying the length of $T_m$ from $4.8\times$RTT down to $1\times$RTT. Then we fix $T_m$ at $1\times$RTT and varying $\eps_{min}$ from $0.01$ to $0.05$. This is not a full exploration of the parameter space, so other settings might actually achieve better trade-off points.} There is a clear convergence speed and stability trade-off: higher $\eps_{min}$ and lower $T_m$ result in faster convergence and higher variance and vice versa. We also show six TCP variants as individual points in the trade-off space. The TCPs either have very long convergence time or high variance. On the other hand, PCC achieves a much better trade-off. For example, PCC with $T_m=1.0\cdot RTT$ and $\eps_{min}=0.02$ achieves the same convergence time and $4.2\times$ smaller rate variance than CUBIC.

Fig.~\ref{tradeoff} also shows the \textbf{benefit of the RCT mechanism} described in \S\ref{sec:controlalgorithm}. While the improvement might look small, it actually helps most in the ``sweet spot'' where convergence time and rate variance are both small, and where improvements are most difficult and most valuable.  Intuitively, with a long monitor interval, PCC gains enough information to make a low-noise decision even in a single interval.  But when it tries to make reasonably quick decisions, multiple RCTs help separate signal from noise. Though RCT doubles the time to make a decision in PCC's Decision State, the convergence time of PCC using RCT only shows slight increase because it makes better decisions. With $T_m=1.0\cdot RTT$ and $\eps_{min}=0.01$, RCT trades $3\%$ increase in convergence time for \textbf{35\% reduction in rate variance}.

\vspace{-3mm}
\subsection{PCC is Deployable}
\vspace{-3mm}
\label{eval:deploy}

\subsubsection{TCP Friendliness}
\label{sec:unfriendly}
\vspace{-2mm}
Opening parallel TCP connections is a very common selfish practice~\cite{wxd, flashget} to overcome TCP's poor performance. We found that PCC is more friendly to normal TCP than $10$ parallel TCP connections. We define \emph{one selfish flow} as either a bundle of $10$ parallel TCP flows, which we call \emph{TCP-Selfish}, or a single PCC flow. We let one normal TCP flow compete with varying number of \emph{selfish flows} on a shared link. For a given number of selfish flows, we compute two average throughputs over $100$s for the normal TCP:  one when it is competing with TCP-Selfish, and one when it is competing with PCC.  We call the ratio between these throughputs the ``relative unfriendliness ratio''. As shown in Figure~\ref{unfriendliness}, under various network conditions, when the number of selfish flows increases, PCC is actually more friendly than TCP-Selfish and therefore, \textbf{PCC is actually safer to use in the commercial Internet than some methods people are already using}.

Moreover, simply using TCP-Selfish cannot achieve consistently high performance and stability like PCC. For example, in the lossy satellite environment, TCP-Selfish will still operate far from optimal and it is hard to decide how many concurrent connections to use to achieve convergence and stability. Moreover, initiating parallel connections involves added overhead in many applications.

\vspace{-4mm}
\subsubsection{Flow Completion Time for Short Flows}
\vspace{-2mm}

Will the ``learning'' nature of PCC harm flow completion time (FCT)? In this section, we resolve this concern by showing that with a startup phase similar to TCP (\S\ref{sec:prototype}), PCC achieves similar FCT for short flows.

We set up a link on Emulab 15 Mbps bandwidth and 60ms RTT.  The sender sends short flows of $100$KB each to receiver. The interval between two short flows is exponentially distributed with mean interval chosen to control the utilization of the link. As shown in Fig.~\ref{shortflowfct}, with network load ranging from $5\%$ to $75\%$, PCC achieves similar FCT at the median and 95th percentile. The 95th percentile FCT with $75\%$ utilization is $20\%$ longer than TCP. However, we believe this is a solvable engineering issue. The purpose of this experiment is to show PCC does not fundamentally harm short flow performance.  There is clearly room for improvement in the startup algorithm of all these protocols, but optimization for fast startup is intentionally outside the scope of this paper because it is a largely separate problem (the same sort of improvements could be applied to either TCP or PCC).

\vspace{-3mm}
\subsection{Alternate Utility Functions with FQ}
\vspace{-3mm}

\label{sec:eval-fq}

In this section, we show a unique feature of PCC: expressing different data transfer objectives by using different utility functions. We only evaluate this feature in a per-flow fair queueing (FQ) environment; with a FIFO queue, the utility function may (or may not) affect dynamics and we leave that to future work. The key fact we want to highlight is that even with Fair Queuing, TCP still needs a complicated active queue management (AQM) mechanism in the network to cater to different applications' objectives as observed in~\cite{nosilverbullet}, while for \name, FQ is sufficient.

\vspace{-5mm}
\subsubsection{Latency Sensitive Applications}
\vspace{-2mm}

As discussed in \S\ref{sec:diffutility}, with a simple resource isolation mechanism such as FQ available in the network, PCC has a feature outside the scope of the TCP family: \name can explicitly express heterogenous data delivery objectives just by plugging in different utility functions, without the need for complex AQM mechanism, which \cite{nosilverbullet} is necessary to cater to different applications' objective.

A key point to support the conclusion in~\cite{nosilverbullet} is that no single AQM is best for all applications.  For example, ``Bufferbloat + FQ + TCP'' is better for throughput-hungry applications but ``Codel + FQ + TCP'' is much better for latency-sensitive interactive applications. In~\cite{nosilverbullet}, an interactive flow is defined as a long-running flow that has the objective of maximizing its throughput-delay ratio, called the \emph{power.} To make our point, we take this scenario and show with PCC, ``Bufferbloat + FQ'' and ``Codel + FQ'' render no power difference for interactive flows and with either, \name achieves higher power than ``TCP + Codel + FQ''.


\begin{figure}[t]
\label{fig:codel}
\centering
{{\includegraphics[width=2.8 in]{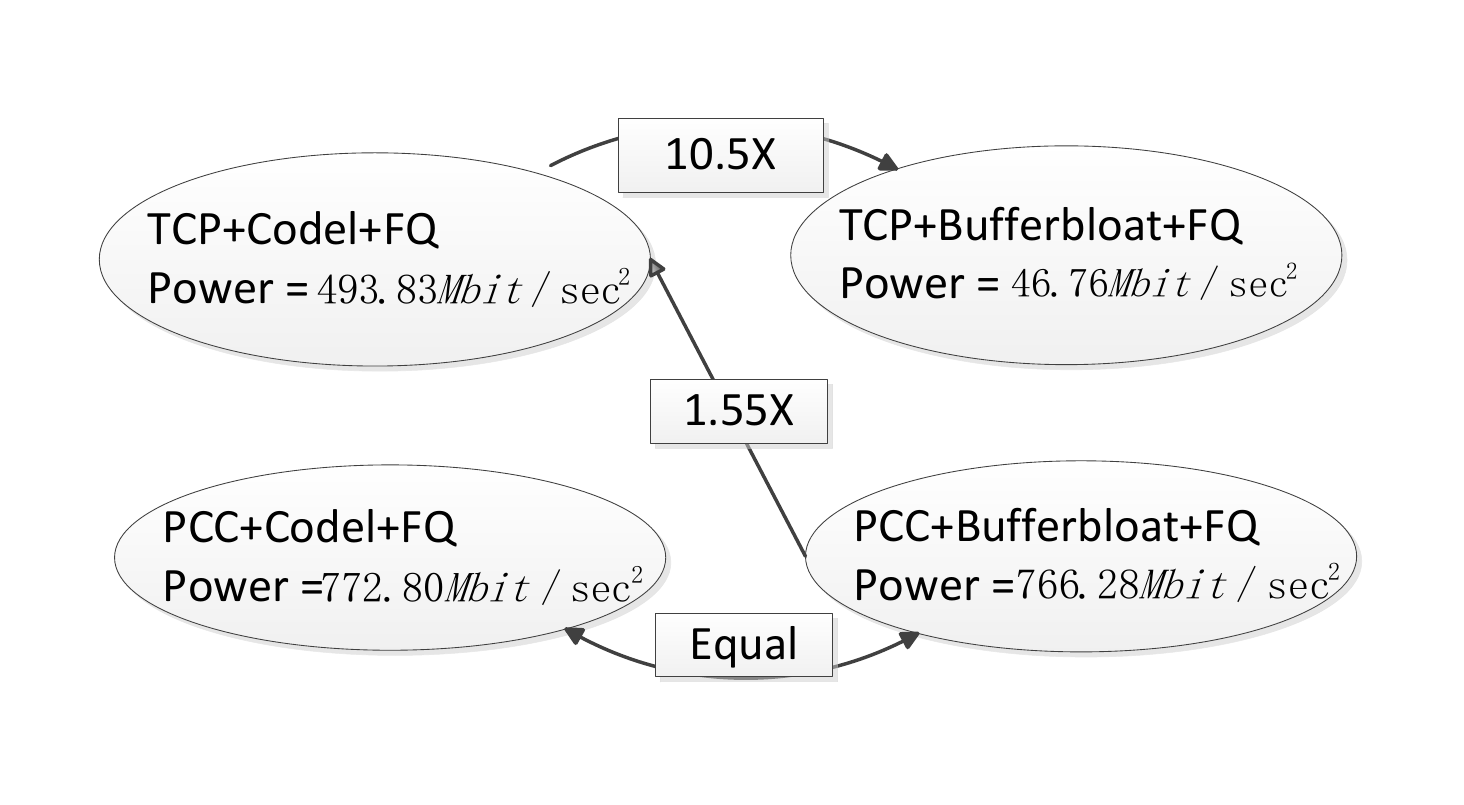}\vspace{-5mm}}
\caption{\small \em \label{codel}
Comparison of Power under different AQM and end-host protocol combination}
}
\vspace{-5mm}
\end{figure}

We set up a transmission pair on Emulab with $40$Mbps bandwidth and $20$ms RTT link running a CoDel implementation~\cite{ubuntucodel} with AQM parameters set as default. We first successfully replicated Sivaraman's simulation result~\cite{nosilverbullet} in a real experiment (Fig.~\ref{codel}): \emph{with TCP CUBIC} and two simultaneous interactive flows, ``FQ + CoDel'' achieves $10.5\times$ more power than ``FQ + Bufferbloat''.

For PCC, we use the following utility function modified from the safe utility function to express the objective of interactive flow:$u_i(x_i)=(T(x_i)\cdot sigmoid(L-0.05)\cdot \frac{RTT_{n-1}}{RTT_{n}}-x_i L)/RTT_n$ where $RTT_{n-1}$ and $RTT_{n}$ is the average RTT of last monitor interval and this monitor interval respectively. This utility function expresses the objective of low latency and avoiding latency increase. With this utility function, we put PCC into the same test setting of TCP. Surprisingly, ``Bufferbloat + FQ'' and ``CoDel + FQ'' achieve essentially the same power for interactive flows. This is because PCC was able to keep buffers very small: we observed \emph{no packet drop} during the experiments even with ``CoDel + FQ'' so PCC's self-inflicted latency never exceed the latency threshold of CoDel. That is to say CoDel becomes useless when PCC is used in end-hosts.

Moreover, ``FQ + Bufferbloat + PCC'' achieves $55\%$ higher power than ``FQ +  Codel + TCP'', indicating that even with AQM, TCP is still bad at expressing the applications' transmission objective.

It is true that in some cases like the LTE environment evaluated in~\cite{nosilverbullet}, AQMs can help end hosts express their objectives better than only do it at end hosts. However, PCC will greatly reduce the need for complicated AQMs in a wide range of network scenarios.

\vspace{-4mm}
\subsubsection{Enduring Excessive Loss}
\vspace{-2mm}

\label{eval:excessiveloss}

Under FIFO queue, to maintain the overall low loss rate, we have to use a ``safe'' utility function (\S\ref{sec:safety}), which also limits the tolerance of random loss. However, with FQ, each sender can strive to achieve its own performance goal without worrying about others. For example, one application can have a highly loss resilient utility function $U=Throughput \cdot (1-Lossrate)$, where its optimal sending rate is its fair share rate and can endure random loss close to $100\%$ in theory. Note that very poor uplink reliability can happen in extreme environments like battlefield or disaster scenarios. We did experiments on Emulab with 100Mbps, 30ms RTT link and forwarding link's loss rate ranging from $10\%$ to $50\%$. PCC's performance is within $97\%$ of the optimal possible achieved throughput even under $50\%$ loss rate and achieves $151\times$ higher throughput than CUBIC under $10\%$ loss rate.






\vspace{-5mm}
\section{Related work}
\vspace{-4mm}
\label{sec:rel}

It has long been clear that TCP lacks enough information, or the right information, to make optimal rate control decisions.  A line of work solved this by using explicit feedback from the network to directly set rate at the end hosts, e.g. XCP~\cite{xcp} and RCP~\cite{rcp}.  But this requires new protocols, router hardware, and packet header formats, so incremental adoption is difficult and lacks the incentive for network operators to deploy.


Numerous designs modify TCP, e.g. \cite{wj+07,CUBIC,ictcp,liu2008tcp,hybla}, but fail to acheive consistent high performance, because they still inherit TCP's hardwired mapping architecture. As we evaluated in \S~\ref{sec:eval}, they only mitigate the problem for the specially assumed network scenarios but still suffer from performance degradation when assumptions are violated. We give another example: FAST TCP~\cite{wj+07} uses prolonged latency as a congestion signal for high BDP connections. However, it models the network queue in an ideal way and its performance degrades under RTT variance~\cite{FASTTCP perf1}, incorrect estimation of baseline RTT~\cite{FASTTCPperf2} and when competing with loss-based TCP protocols.


Remy~\cite{remy, learnabilitytcp} pushes TCP's architecture to extreme: it exhaustively searches through a large number of \emph{hardwired mappings} under a network model with assumed parameters, e.g. number of senders, link speed, etc., and finds the best protocol under that sceanrio. However, like all TCP variants, when the real network deviates from Remy's input assumption, performance degrades~\cite{learnabilitytcp}. Moreover, the real network can have many more ``parameters'' than are in Remy's network model and the result of that is unclear.

Other works, such as PCP~\cite{pcp}, ($4.58\times$ worse than PCC in \S~\ref{eval:planet}), and Packet Pair Flow Control~\cite{keshavpp} utilize techniques like packet-train~\cite{pathload} to probe available bandwidth in the network. However, bandwidth probing (BP) protocols do not observe real performance like PCC does and make unreliable assumptions about the network. For example, real networks can easily violate the assumptions about packet inter-arrival latency embedded in BP (e.g. latency jitter and variation due to middle boxes, software routers or virtualization layers), rendering incorrect estimates that harm performance. When we set up a clean link on emulab with $100$Mbps bandwidth and 30ms RTT with 75KB buffer, PCP~\cite{pcpimplementation} continuously wrongly estimates the available bandwidth as $50-60$Mbps.


Decongestion Control~\cite{raghavan2006decongestion} sends at full line rate, masking loss with erasure coding.  PCC is selfish, but optimizes a utility function and converges to an efficient equilibrium.

Finally, none of the aforementioned work allows the possibility of expressing different sending objectives by plugging in different utility functions as PCC does.


\vspace{-5mm}
\section{Conclusion and Future Work}
\label{sec:conclusion}
\vspace{-4mm}
This paper proposes performance-oriented congestion control architecture and shows its promising potential towards a consistent high performance congestion control architecture. Particularly interesting questions that remain include designing a better learning algorithm, further analysis and experiments on different utility functions' interaction under FIFO queuing, and real-world deployment.


{\bibliographystyle{abbrv}

\bibliography{paper}

\begin{thebibliography}{10}

\bibitem{ubuntucodel}
{Codel Linux implementation}.
\newblock \url{http://goo.gl/06VQqG}.

\bibitem{esnet}
{ESNet}.
\newblock \url{http://www.es.net/}.

\bibitem{flashget}
{FlashGet}.
\newblock \url{flashget.com}.

\bibitem{pccfullproof}
{Full proof of Theorem 1}.
\newblock \url{http://web.engr.illinois.edu/~modong2/full_proof.pdf}.

\bibitem{genitestbed}
{GENI testbed}.
\newblock \url{http://www.geni.net/}.

\bibitem{internet2ion}
{Internet2 ION service}.
\newblock \url{http://webdev0.internet2.edu/ion/}.

\bibitem{level3}
{Level 3 Bandwidth Optimizer}.
\newblock \url{http://goo.gl/KFQ3aS}.

\bibitem{limelight}
{Limelight Orchestrate(TM) Content Delivery}.
\newblock \url{http://goo.gl/M5oHnV}.

\bibitem{pcpimplementation}
{PCP user-spamce implementation}.
\newblock \url{http://homes.cs.washington.edu/~arvind/pcp/pcp-ulevel.tar.gz}.

\bibitem{vsatlatency}
{Satellite link latency}.
\newblock \url{http://goo.gl/xnqCih}.

\bibitem{tellitec}
{TelliShape per-user traffic shaper}.
\newblock \url{http://tinyurl.com/pm6hqrh}.

\bibitem{udt}
{UDT: UDP-based data transfer}.
\newblock \url{udt.sourceforge.net}.

\bibitem{wxd}
{wxDownload Fast}.
\newblock \url{dfast.sourceforge.net}.

\bibitem{planetlab}
{PlanetLab} | {An} open platform for developing, deploying, and accessing
  planetary-scale services.
\newblock July 2010.
\newblock \url{http://www.planet-lab.org}.

\bibitem{dctcp}
M.~Alizadeh, A.~Greenberg, D.~Maltz, J.~Padhye, P.~Patel, B.~Prabhakar,
  S.~Sengupta, and M.~Sridharan.
\newblock Data center tcp ({DCTCP}).
\newblock {\em Proc. ACM SIGCOMM}, May 2010.

\bibitem{pcp}
T.~Anderson, A.~Collins, A.~Krishnamurthy, and J.~Zahorjan.
\newblock Pcp: Efficient endpoint congestion control.
\newblock In {\em Proc. NSDI}, 2006.

\bibitem{appenzeller2004sizing}
G.~Appenzeller, I.~Keslassy, and N.~McKeown.
\newblock {\em Sizing router buffers}.
\newblock 2004.

\bibitem{FASTTCPperf1}
H.~Bullot and R.~L. Cottrell.
\newblock Evaluation of advanced tcp stacks on fast longdistance production
  networks.
\newblock {\em Proc. PFLDNeT}, February 2004.

\bibitem{rcp}
M.~Caesar, D.~Caldwell, N.~Feamster, J.~Rexford, A.~Shaikh, and K.~van~der
  Merwe.
\newblock Design and implementation of a routing control platform.
\newblock {\em Proc. NSDI}, April 2005.

\bibitem{hybla}
C.~Caini and R.~Firrincieli.
\newblock {TCP Hybla: a TCP enhancement for heterogeneous networks}.
\newblock {\em International Journal of Satellite Communications and
  Networking}, 2004.

\bibitem{understandingincast}
Y.~Chen, R.~Griffith, J.~Liu, R.~H. Katz, and A.~D. Joseph.
\newblock Understanding tcp incast throughput collapse in datacenter networks.
\newblock {\em Proc. ACM SIGCOMM Workshop on Research on Enterprise
  Networking}, 2009.

\bibitem{oscars}
{ESnet}.
\newblock {Virtual Circuits (OSCARS)}, May 2013.
\newblock \url{http://goo.gl/qKVOnS}.

\bibitem{bufferbloat}
J.~Gettys and K.~Nichols.
\newblock Bufferbloat: Dark buffers in the internet.
\newblock December 2011.

\bibitem{splendidisolation}
S.~Gutz, A.~Story, C.~Schlesinger, and N.~Foster.
\newblock Splendid isolation: a slice abstraction for software-defined
  networks.
\newblock {\em Proc. ACM SIGCOMM Workshop on Hot Topics in Software Defined
  Networking}, 2012.

\bibitem{CUBIC}
S.~Ha, I.~Rhee, and L.~Xu.
\newblock {CUBIC: a new TCP-friendly high-speed TCP variant}.
\newblock {\em Proc. ACM SIGOPS}, 2008.

\bibitem{satellitetutorial}
Y.~Hu and V.~Li.
\newblock Satellite-based internet: a tutorial.
\newblock {\em Communications Magazine}, 2001.

\bibitem{pathload}
M.~Jain and C.~Dovrolis.
\newblock Pathload: A measurement tool for end-to-end available bandwidth.
\newblock In {\em Proc. Passive and Active Measurement (PAM)}, 2002.

\bibitem{jiang2012improving}
J.~Jiang, V.~Sekar, and H.~Zhang.
\newblock Improving fairness, efficiency, and stability in http-based adaptive
  video streaming with festive.
\newblock In {\em Proc. CoNEXT}, 2012.

\bibitem{xcp}
D.~Katabi, M.~Handley, and C.~Rohrs.
\newblock Congestion control for high bandwidth-delay product networks.
\newblock {\em Proc. ACM SIGCOMM}, August 2002.

\bibitem{keshavpp}
S.~Keshav.
\newblock {\em The packet pair flow control protocol}.
\newblock ICSI, 1991.

\bibitem{liu2008tcp}
S.~Liu, T.~Ba{\c{s}}ar, and R.~Srikant.
\newblock Tcp-illinois: A loss-and delay-based congestion control algorithm for
  high-speed networks.
\newblock {\em Performance Evaluation}, 65(6):417--440, 2008.

\bibitem{westwood}
S.~Mascolo, C.~Casetti, M.~Gerla, M.~Sanadidi, and R.~Wang.
\newblock Tcp westwood: Bandwidth estimation for enhanced transport over
  wireless links.
\newblock 2001.

\bibitem{winds}
H.~Obata, K.~Tamehiro, and K.~Ishida.
\newblock Experimental evaluation of tcp-star for satellite internet over
  winds.
\newblock 2011.

\bibitem{faircloud}
L.~Popa, G.~Kumar, M.~Chowdhury, A.~Krishnamurthy, S.~Ratnasamy, and I.~Stoica.
\newblock Faircloud: Sharing the network in cloud computing.
\newblock {\em Proc. ACM SIGCOMM}, August 2012.

\bibitem{elasticswitch}
L.~Popa, P.~Yalagandula, S.~Banerjee, J.~C. Mogul, Y.~Turner, and J.~R. Santos.
\newblock Elasticswitch: Practical work-conserving bandwidth guarantees for
  cloud computing.
\newblock {\em Proc. ACM SIGCOMM}, August 2013.

\bibitem{prasad2007router}
R.~Prasad, C.~Dovrolis, and M.~Thottan.
\newblock Router buffer sizing revisited: the role of the output/input capacity
  ratio.
\newblock 2007.

\bibitem{raghavan2006decongestion}
B.~Raghavan and A.~Snoeren.
\newblock Decongestion control.
\newblock {\em Proc. HotNets}, 2006.

\bibitem{Rosen1965}
J.~Rosen.
\newblock Existence and uniqueness of equilibrium point for concave n-person
  games.
\newblock {\em Econometrica}, 1965.

\bibitem{nosilverbullet}
A.~Sivaraman, K.~Winstein, S.~Subramanian, and H.~Balakrishnan.
\newblock {No silver bullet: extending SDN to the data plane}.
\newblock {\em Proc. HotNets}, 2013.

\bibitem{learnabilitytcp}
A.~Sivaraman, K.~Winstein, P.~Thaker, and H.~Balakrishnan.
\newblock An experimental study of the learnability of congestion control.
\newblock 2014.

\bibitem{FASTTCPperf2}
L.~Tan, C.~Yuan, and M.~Zukerman.
\newblock {FAST TCP}: Fairness and queuing issues.
\newblock {\em IEEE Communication Letter}, August 2005.

\bibitem{vsat}
{VSAT Systems}.
\newblock {TCP/IP protocol and other applications over satellite}.
\newblock \url{http://goo.gl/E6q6Yf}.

\bibitem{wj+07}
D.~Wei, C.~Jin, S.~Low, and S.~Hegde.
\newblock {FAST TCP}.
\newblock {\em IEEE/ACM Trans. Networking}, December 2006.

\bibitem{emulab}
B.~White, J.~Lepreau, L.~Stoller, R.~Ricci, G.~Guruprasad, M.~Newbold,
  M.~Hibler, C.~Barb, and A.~Joglekar.
\newblock An integrated experimental environment for distributed systems and
  networks.
\newblock {\em Proc. OSDI}, December 2002.

\bibitem{remy}
K.~Winstein and H.~Balakrishnan.
\newblock Tcp ex machina: Computer-generated congestion control.
\newblock {\em Proc. ACM SIGCOMM}, 2013.

\bibitem{ictcp}
H.~Wu, Z.~Feng, C.~Guo, and Y.~Zhang.
\newblock {ICTCP}: Incast congestion control for {TCP} in data center networks.
\newblock {\em Proc. CoNEXT}, 2010.

\end{thebibliography}
}

\end{document}